\documentclass{article}
\usepackage[T1]{fontenc}
\usepackage{graphicx,fancyhdr,tabularx,amsmath,amssymb,color}
\usepackage{subcaption}
\usepackage{physics}
\usepackage{breqn}
\usepackage[export]{adjustbox}
\usepackage[shortlabels]{enumitem}
\usepackage{url}
\usepackage{float}
\usepackage{svg}
\usepackage{titlesec, blindtext, color}
\definecolor{gray75}{gray}{0.75}
\usepackage{verbatim}
\usepackage{multirow}
\usepackage{titlesec}
\usepackage{todonotes}
\usepackage{multicol}
\usepackage{wrapfig}
\usepackage{sidecap} 

\NewDocumentCommand{\evalat}{sO{\big}mm}{%
  \IfBooleanTF{#1}
   {\mleft. #3 \mright|_{#4}}
   {#3#2|_{#4}}%
}
\newcommand{\figWidthProfiles}{0.45\textwidth}

\usepackage[
backend=biber,
sorting=none,
defernumbers=true,
giveninits=true,
maxbibnames=10,
doi=false,
isbn=false,
url=false,
eprint=false,
arxiv=false,
natbib=true,
date=year]{biblatex}
\bibliography{literature.bib}
\makeatletter 
\let\blx@rerun@biber\relax
\makeatother

\usepackage{listings}
\usepackage{xcolor}

\usepackage{authblk} 

\definecolor{codegreen}{rgb}{0.2,0.7,0.2}
\definecolor{codeblue}{rgb}{0,0.2,0.5}
\definecolor{codered}{rgb}{0.7,0.3,0.3}
\definecolor{backcolour}{rgb}{0.95,0.95,0.92}
\lstdefinestyle{mystyle}{
    backgroundcolor=\color{backcolour},   
    commentstyle=\color{codeblue},
    keywordstyle=\color{codegreen},
    stringstyle=\color{codered},
    basicstyle=\ttfamily\footnotesize,
    breakatwhitespace=false,         
    breaklines=true,                 
    captionpos=b,                    
    keepspaces=true,                 
    numbers=left,                    
    numbersep=5pt,                  
    showspaces=false,                
    showstringspaces=false,
    showtabs=false,                  
    tabsize=2
}
\newcommand{\vv}[1]{\boldsymbol{#1}}

\usepackage[a4paper, total={6.3in, 9in}]{geometry}

\begin{document}
\title{Chiral flows can induce neck formation in viscoelastic surfaces}
\author[1,2]{E. M. de Kinkelder}
\author[3,4,5]{E. Fischer-Friedrich \thanks{elisabeth.fischer-friedrich@tu-dresden.de}}
\author[1,2]{S. Aland \thanks{sebastian.aland@math.tu-freiberg.de}}
\affil[1]{Institute of Numerical Analysis and Optimization, Technische Universtit\"at Bergakademie Freiberg, Freiberg, Germany}
\affil[2]{Faculty of Informatics/Mathematics, Hochschule f\"ur Technik und Wirtschaft, Dresden, Germany}
\affil[3]{Cluster of Excellence Physics of Life, Technische Universit\"at Dresden, Dresden, Germany}
\affil[4]{Biotechnology Center, Technische Universit\"at Dresden, Dresden, Germany}
\affil[5]{Faculty of Physics, Technische Universit\"at Dresden, Dresden, Germany}
\date{\small\today}

\maketitle
\begin{abstract}
    During division in animal cells, the actomyosin cortex has been found to exhibit  counter-rotating cortical flows, also known as chiral flows, along the axis of division. 
    Furthermore, such chiral surface flows were shown to influence cellular rearrangements  and drive the left-right symmetry breaking in developing organisms. 
    In spite of this prospective biological importance, at the current state, no numerical simulations have been done to study the influence of chiral flows on the  cell cortex shape. To deepen the insight on that matter, we present here a numerical study of an axi-symmetric viscoelastic surface embedded in a viscous fluid. To investigate the influence of a chiral flow field on the surface shape and material transport, we impose a generic counter-rotating force field on this surface which induces a chiral flow field.  Notably, we find that the building of a neck, as is observed during cell division, occurs if there is a strong shear elastic component. Furthermore we find that a large areal relaxation time results in flows towards the equator of the surface. These flows assist the transport of a surface concentration during the forming of a contractile ring. Accordingly, we show that chiral forces by themselves can drive pattern formation and stabilise contractile rings at the equator.
\end{abstract}



\section{Introduction}
Most organisms are chiral, i.e. they exhibit left-right asymmetries and are not superimposable on their mirror images \cite{Lebreton2018,Wood1997,Naganathan2016,Chen2012,Tee2015,Vandenberg2013}. 
In particular, chiral flows at the level of the fertilised egg were shown to be linked to symmetry breaking and the establishment of the left-right body axis in several invertebrate species \cite{Danilchik2006, Naganathan2014, Blum2018}.
At the level of the cell, shape is mainly regulated by the actomyosin cortex, a thin biopolymer network at the surface of the cell right underneath the plasma membrane \cite{Salbreux2012}.
Notably, chiral asymmetries in the flow field of the cortex were apparent during cell division of some egg cells \cite{Pimpale2020,Danilchik2006,Blum2018}. However, at the current state, the influence of chiral cortical flows on cellular morphogenesis remains elusive.

In order to better understand the influence of chiral cortical flows on cellular shape evolution, we study a numerical model of an active viscoelastic surface subject to an induced azimuthal flow field whose velocities are of opposite direction on the left and right hemisphere. In the following, we will refer to these azimuthal flows as chiral flows. Such flows have been observed in the cell during cytokinesis \cite{Naganathan2016,Pimpale2020}. The source of these chiral flows are most likely forces generated by the cytoskeleton. Local torques, caused by the helix structure of the actin filaments could result in large scale chiral flows. These, in turn, could influence the dynamics during cell division \cite{Sase1997,Ali2002,Mizuno2011,Naganathan2014}. To see this, consider balloon animals as an analogy. To make a balloon animal, you have to create a neck in the balloon; this is done by twisting the balloon in opposite directions, which will result in a neck between your hands. This illustrates that a chiral force field can induce a neck in an elastic surface.
Most numerical simulations of the cell cortex assume it to be an elastic shell or purely viscous \cite{Stadtlander2013, Mokbel_ACS_2017}. However, experimental measurements indicate that the cell cortex is viscoelastic with timescale-dependent mechanical properties \cite{Fischer-Friedrich2016, Hosseini2021,Bonfanti2020,Khalilgharibi2019,Mokbel2020}. In particular, it has been shown that the cortex is stiff and dominantly elastic at short time scales and has fluid like properties at long time scales \cite{Fischer-Friedrich2016,Hosseini2021, Bonfanti2020}, a behaviour which in simplest form can be described by a Maxwell model. In this paper, we use the upper-convected-surface Maxwell model as described in \cite{DeKinkelder2021} to model the viscoelastic surface.
This type of viscoelasticity can be best illustrated by considering the 1D Maxwell element in Fig. \ref{fig:domain} left. Here, a viscous dashpot with viscosity $\eta$ is in series with an elastic spring with elastic modulus $G$. When the Maxwell element is dilated, first, elastic stresses are stored in the spring but then dissipated by viscous dissipation through the sliding of the dashpot. The ratio $\eta/G$ defines a time scale $\tau$ of stress relaxation. 

Previous research has investigated chiral flows and forces experimentally \cite{Pimpale2020, Naganathan2016} or has taken a theoretical approach \cite{Furthauer2012,Furthauer2013}. But as numerical models for deforming viscoelastic fluid surfaces have only been developed recently \cite{DeKinkelder2021,jaensson2021computational}, no computational model study has been performed combining viscoelastic surfaces and chiral flows. 
Accordingly, the effect of chiral forces on the flows and shape dynamics of a deformable surface are completely unexplored so far. 
The goal of this paper is to study the influence of the viscoelastic parameters on the emergent shape of the surface and the distribution of a surface  concentration field under the influence of chiral forces. 

In Sec. \ref{sec:Governing equations}, we introduce the model. In Sec. \ref{sec:Results}, we discuss the numerical experiments that have been done and corresponding results. In particular, we study the influence of the viscoelastic parameters on the shape in Secs. \ref{sec:viscoelastic}-\ref{sec:relaxationTimes}. Then, we show the potential of the chiral force field to stabilise a contractile ring in an active surface during cell division in Sec.~\ref{sec:contractileRing}. These results are then summarised and put into a biological perspective in Sec. \ref{sec:discussion}.

\section{Governing equations}
\label{sec:Governing equations}
\begin{figure}[H]
\centering
    \includegraphics[width = 0.49\textwidth]{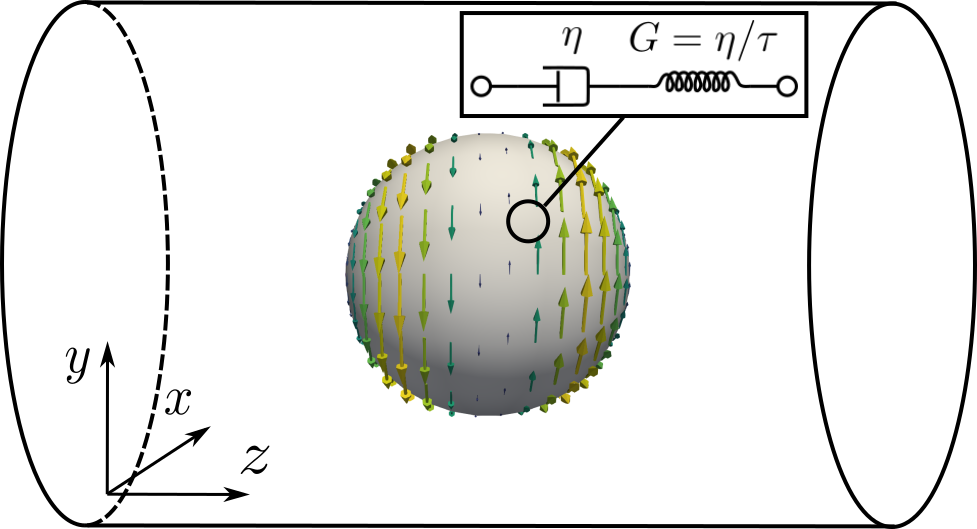} 
    \includegraphics[width = 0.49\textwidth]{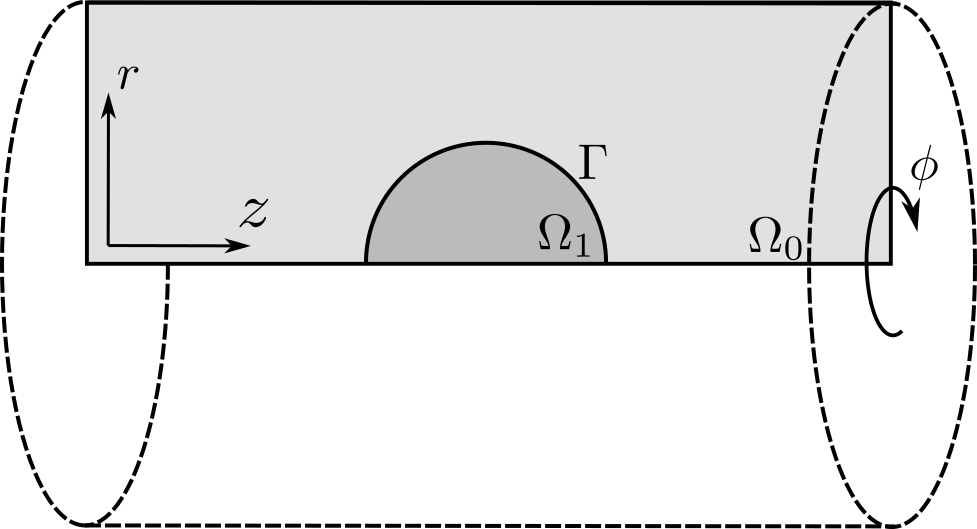}
    \caption{\textbf{Left:} Sketch of the 3D domain: a viscoelastic surface is embedded in surrounding fluids and subject to a counter-rotating (chiral) force field.  The schematic of the Maxwell model illustrates the material nature of the viscoelastic surface with surface viscosity $\eta$, elastic modulus $G$ and relaxation time $\tau$. In our model however, we consider two parts of viscoelasticity corresponding to areal and shear stresses. \textbf{Right:} Sketch of the axi-symmetric domain and axes ($z$ longitudinal, $r$ radial, $\phi$ azimuthal). The internal and external fluids are named $\Omega_1$ and $\Omega_0$, the surface is named $\Gamma$. Exploiting axi-symmetric conditions, computations can be reduced to the 2D shaded domain.}
    \label{fig:domain}
\end{figure}
We model a viscoelastic cell surface embedded in viscous fluids. A sketch of the domain is given in Fig. \ref{fig:domain} left. The domains are labelled $\Omega_0$ for the external fluid and $\Omega_1$ for the internal fluid. The cell surface $\Gamma$ separates the fluids and is assumed to have zero thickness as we anticipate that the actin cortex is thin as compared to the cell radius \cite{Clark2013}. Although we use an axisymmetric implementation (Fig. \ref{fig:domain} right), the governing equations are given in three dimensions in the following.

Given the small length scale of biological cells, we anticipate low Reynolds numbers and model the surrounding and cytoplasmic fluid as incompressible Stokes fluids,
\begin{align}
    \nabla \cdot \vv{v} &= 0 & \text{in } \Omega_0 \cup \Omega_1,\\
    0 &= -\nabla p + \eta_i \nabla \cdot (\nabla \vv{v} + (\nabla \vv{v})^T) & \text{in } \Omega_0 \cup \Omega_1.
\end{align}
Here $\vv{v}$ represents the velocity, $p$ the pressure and $\eta_i$ the fluid viscosity in $\Omega_i$.

The enclosing surface $\Gamma$ is viscoelastic and the corresponding viscoelastic stress $S$ is decomposed into its areal part $\tr S$ and shear part $\bar{S}$, such that $S = \bar{S} + \frac{1}{2} (\tr S) P$. The matrix  $P = I - \vv{n} \otimes \vv{n}$ is the projection matrix, it projects a vector on the surface with surface normal vector $\vv{n}$. 
In a 2D surface, mechanical resistance with regard to shear deformation and area dilation needs to be described by a set of two elastic moduli and corresponding viscosities in the case of a Maxwell-type viscoelasticity. Equivalently, these parameters can be expressed by two viscosities and two relaxation time scales. Therefore, we will use shear and areal viscosities $\eta_S$ and $\eta_A$ and the shear and areal relaxation times $\tau_S$ and $\tau_A$ as mechanical parameters of the surface. 
The evolution of the stress is determined by changes in surface morphology and surface flows. Consistent evolution equations for the stress components were derived in \cite{DeKinkelder2021} as
\begin{align}
    \bar{S} &= 2 \eta_S \bar{D} - \tau_S \overline{\overset{\nabla}{\delta} \bar{S}}  &\text{on } \Gamma, \label{eq:devS}\\
    \tr(S) &= 2 \eta_A \tr(D) + \tau_A \left(2(\bar{S}:\nabla_\Gamma \vv{v}) + \tr (S)\tr (D) -\partial_{t}^\bullet \tr (S) \right) &\text{on } \Gamma. \label{eq:trS}
\end{align}
Here, the surface rate of deformation tensor is defined as $D = P\left(\nabla_\Gamma\vv{v}+\nabla_\Gamma\vv{v}^T\right)P$. Its traceless part is $\bar{D}=D-\frac{1}{2}(\tr D)P$. 
The operators are the material derivative $\partial_{t}^\bullet$, and the traceless upper convected surface derivative, which is defined as 
\begin{equation}
    \overline{\overset{\nabla}{\delta}\bar{S}} = \partial_{t}^\bullet \bar{S} - \nabla_\Gamma \vv{v} \bar{S} - \bar{S} (\nabla_\Gamma \vv{v})^T + P(\bar{S} : \nabla_\Gamma \vv{v}) - \tr (S) \bar{D}. \label{eq:objSurfDeriv}
\end{equation}
for a traceless tensor $\bar{S}$. 

We also track the evolution of a surface quantity (e.g. surface-bound protein like myosin). The dynamics of such a surface concentration $c$ is given by the advection diffusion Eq. \cite{Lucas_paper}
\begin{equation}
    \partial_{t}^\bullet c + c(\nabla_\Gamma \cdot \vv{v}) = D_c \Delta_\Gamma c,
\end{equation}
where $D_c$ is the diffusion coefficient. 
If the surface quantity represents molecular motor proteins, the corresponding force field can be included as an isotropic active surface tension stress given by 
\begin{equation}
     S_a =  \xi f(c) P, 
     \label{eq:active tension}
\end{equation}
where the parameter $\xi$ regulates the strength of the active contribution,  $f(c)$ which is expressed  consistently with previous literature\cite{Mietke2019, Mietke2019a, Lucas_paper} by a monotonically increasing Hill function $f=\frac{c^2}{c^2+c_0^2}$, where $c_0$ is the constant equilibrium concentration on the surface $\Gamma$ \cite{Mietke2019, Lucas_paper, Bonati2022}. 
Note, that for most parts of this article, the parameter $\xi$ will be set to $0$ such that there is no feedback from $c$ on the mechanics of the system. This is to remove the interplay between deformation of the surface caused by the chiral forces and the deformation caused by the active surface tension.

Finally, to systematically analyse the effect of chiral forces we prescribe a well-defined generic counter rotating force field
\begin{equation}
    \vv{f}_c = \alpha \hat{\vv{e}}_\phi r_0 z_0.
    \label{eq:chiralForce}
\end{equation}
Here, $\hat{\vv{e}}_\phi$ is the base vector in the rotational direction, $r$ is the distance from the axis of rotation and $z$ is the location along the axis of rotation, see Fig.\ref{fig:domain} right. The surface is centred at the origin $r=z=0$. The values $r_0$ and $z_0$ are the initial  $r$ and $z$ coordinates of the material points on the surface. The strength of the chiral force field is scaled by the factor $\alpha$. 
In Fig. \ref{fig:domain} left, a schematic representation of the chiral force field is given.
Putting together the viscoelastic, active, chiral and fluid forces on the surface, results in the boundary condition
\begin{equation}
    \nabla_\Gamma \cdot S + \nabla_\Gamma \cdot S_a + \vv{f}_c = \left[-p{I} + \eta_i (\nabla \vv{v} + (\nabla \vv{v})^T) \right]_0^1\vv{n} \qquad \text{on } \Gamma,
    \label{eq:jumpCondition}
\end{equation}
where the square brackets denote the discontinuous jump of the enclosed tensor across the surface.

\subsection{Non-dimensional equations and parameters}
The equations are non-dimensionalised using the initial cell radius $R$ 
as characteristic length scale and setting the characteristic timescale $t_{sc} = \frac{\eta_1}{\alpha R^2}$. This can be interpreted as the resistance of the fluid divided by the strength of the chiral force field. 
The system of equations in dimensionless form is
\begin{align}
    \bar{S} &= 2 \bar{D} - \hat{\tau}_S \left( \partial_{t}^\bullet \bar{S} - \nabla_\Gamma \vv{v} \bar{S} - \bar{S} (\nabla_\Gamma \vv{v})^T + P(\bar{S} : \nabla_\Gamma \vv{v}) - \frac{\hat{\eta}_A}{\hat{\eta}_S} \tr (S) \bar{D} \right) & \text{on } \Gamma, \label{eq:devSscaled} \\
    \tr(S) &= 2 \tr (D) + \hat{\tau}_A \left(\frac{2\hat{\eta}_S}{\hat{\eta}_A}(\bar{S}:\nabla_\Gamma \vv{v}) + \tr (S)\tr (D) -\partial_{t}^\bullet \tr S \right) & \text{on } \Gamma, \label{eq:trSscaled}\\
    0 &= \nabla \cdot \vv{v} & \text{in } \Omega_0 \cup \Omega_1, \label{eq:stokes1scaled} \\
    0 &= -\nabla p + 2 \frac{\eta_i}{\eta_1} \nabla \cdot \left(\nabla\vv{v}+\nabla\vv{v}^T\right) & \text{in } \Omega_0 \cup \Omega_1, \label{eq:stokes2scaled} \\
     0 &=\left[ -p{I} + 2 \frac{\eta_i}{\eta_1}\left(\nabla\vv{v}+\nabla\vv{v}^T\right)\right]_0^1{\bf n} - \nabla_\Gamma \cdot \left(S + S_a \right) - \hat{\vv{e}}_\phi r_0 z_0 & \text{on } \Gamma, \label{eq:jumpConditionScaled} \\
    S &= \frac{\hat{\eta}_A}{2} \tr(S) P + \hat{\eta}_S \bar{S} & \text{on } \Gamma, \\
     \hat{D}_c \Delta_\Gamma c &= \partial_{t}^\bullet c + c(\nabla_\Gamma \cdot \vv{v})& \text{on } \Gamma, \label{eq:concentrationScaled} \\
    S_a &= \hat{\xi} f(c) P & \text{on } \Gamma. \label{eq:activeSurfTenScaled}
\end{align}
We use the dimensionless parameters given in Table \ref{tab:params}.

\begin{table}[H]
\centering
\begin{tabular}{l|l|l}
Quantity & Symbol & Range \\\hline
Scaled relaxation time               & $\hat{\tau}_{A/S} = \frac{\tau_{A/S} \alpha R^2}{\eta_1}$ &$[10^{-2}, 10^3]$  \\
Scaled surface viscosity             & $\hat{\eta}_{A/S} = \frac{\eta_{A/S}}{\eta_1 R}$  & $[10^{-1}, 10^4]$  \\
Scaled elastic modulus               & $\hat{G}_{A/S} = \frac{\hat{\eta}_{A/S}}{\hat{\tau}_{A/S}}$    & $[10^{-4}, 10^2]$  \\
Fluid viscosity ratio & $\frac{\eta_i}{\eta_1}$, $i \in \{0,1\}$ & 1 \\
Scaled diffusion coefficient         & $\hat{D}_c = \frac{D_c \eta_1}{\alpha R^4}$  & $10^{-4}$  \\
Activity to chiral force field ratio &  $\hat{\xi} = \frac{\xi}{R^3 \alpha}$  & $0$ \\\hline
\end{tabular}
\caption{Definitions and values of the non-dimensional parameters. In our simulations, unless stated otherwise, the parameters will be chosen as they are in this table.}
\label{tab:params}
\end{table}

\section{Results}
\label{sec:Results}
To study the influence of the parameters on the mechanics, we simulated the model using the discretization described in Appendix \ref{app:numerical implementation}. We varied the viscoelastic non-dimensional parameters $\hat{\tau}_A$, $\hat{\tau}_S$, $\hat{\eta}_A$, $\hat{\eta}_S$, $\hat{G}_A$ and $\hat{G}_S$ to observe their influence on the surface shape dynamics. Unless stated otherwise, simulations start with a spherical surface $\Gamma$ with dimensionless radius 1. The initial stress $S$ and velocity $\vv{v}$ are set to zero, the dimensionless concentration $c$ is initially equal to one.

\begin{figure}[H]
    \centering
    \begin{subfigure}{0.49\textwidth}
        \centering
        \includegraphics[width=0.99\textwidth]{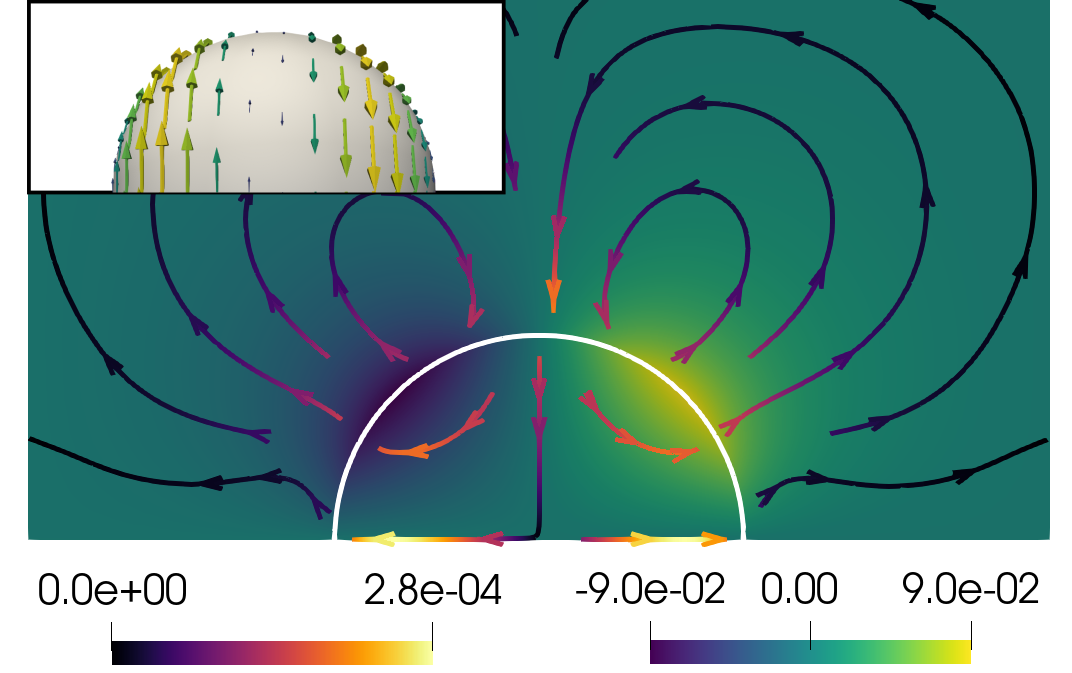}
        \caption{$t=1$}
        \label{fig:example1}
    \end{subfigure}
    \begin{subfigure}{0.49\textwidth}
        \centering
        \includegraphics[width=0.99\textwidth]{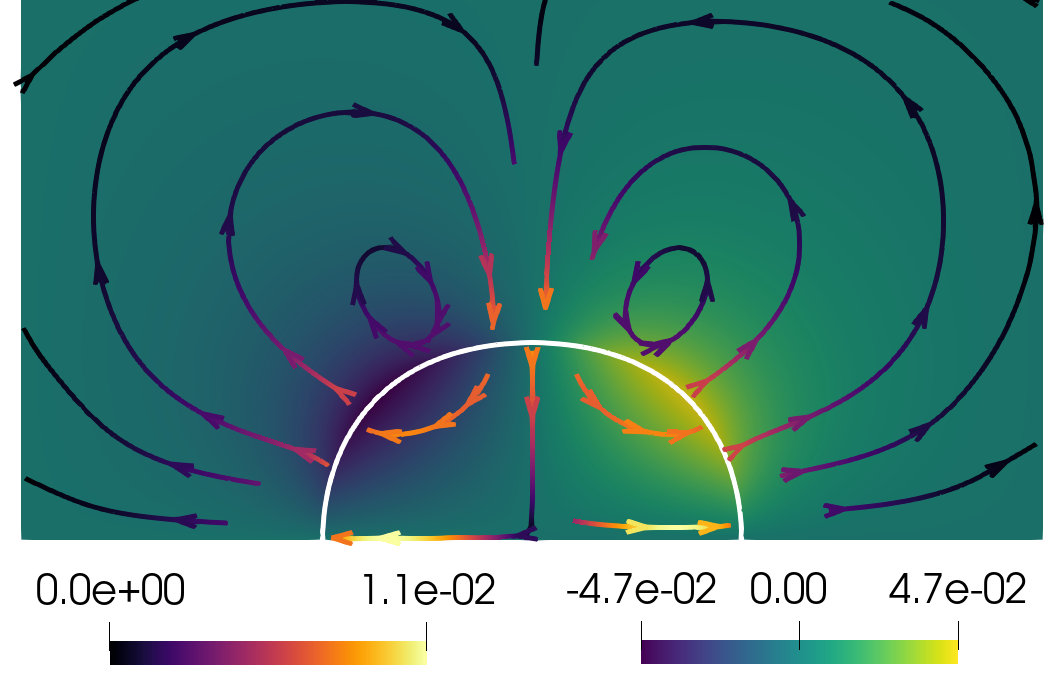}
        \caption{$t=10$}
        \label{fig:example10}
    \end{subfigure}
    \begin{subfigure}{0.49\textwidth}
        \centering
        \includegraphics[width=0.99\textwidth]{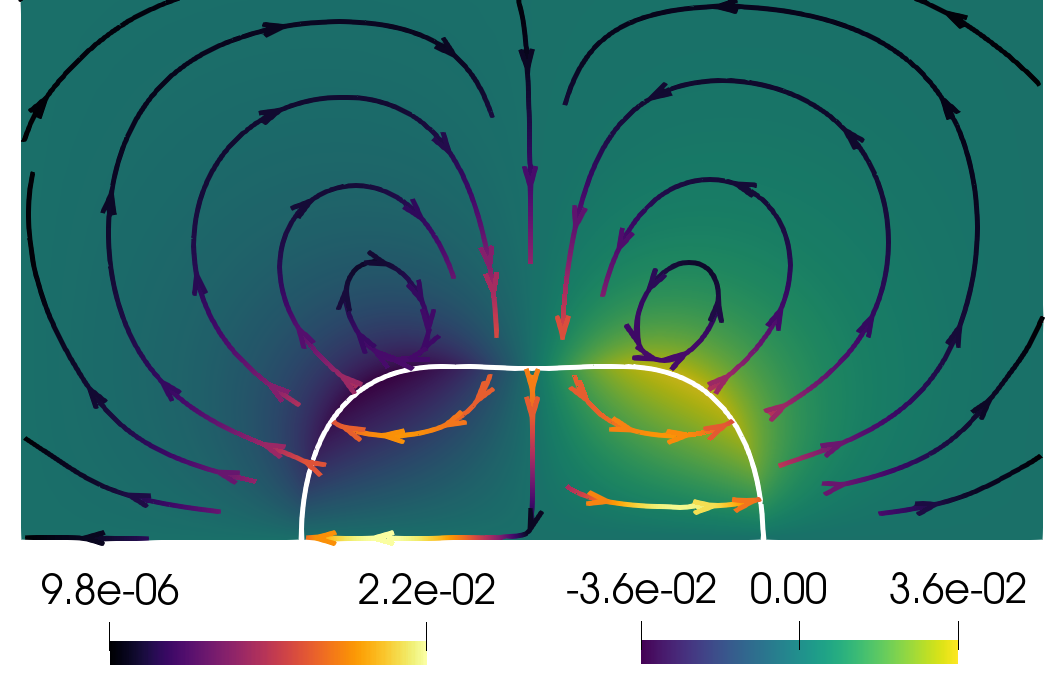}
        \caption{$t=20$}
        \label{fig:example20}
    \end{subfigure}
    \begin{subfigure}{0.49\textwidth}
        \centering
        \includegraphics[width=0.99\textwidth]{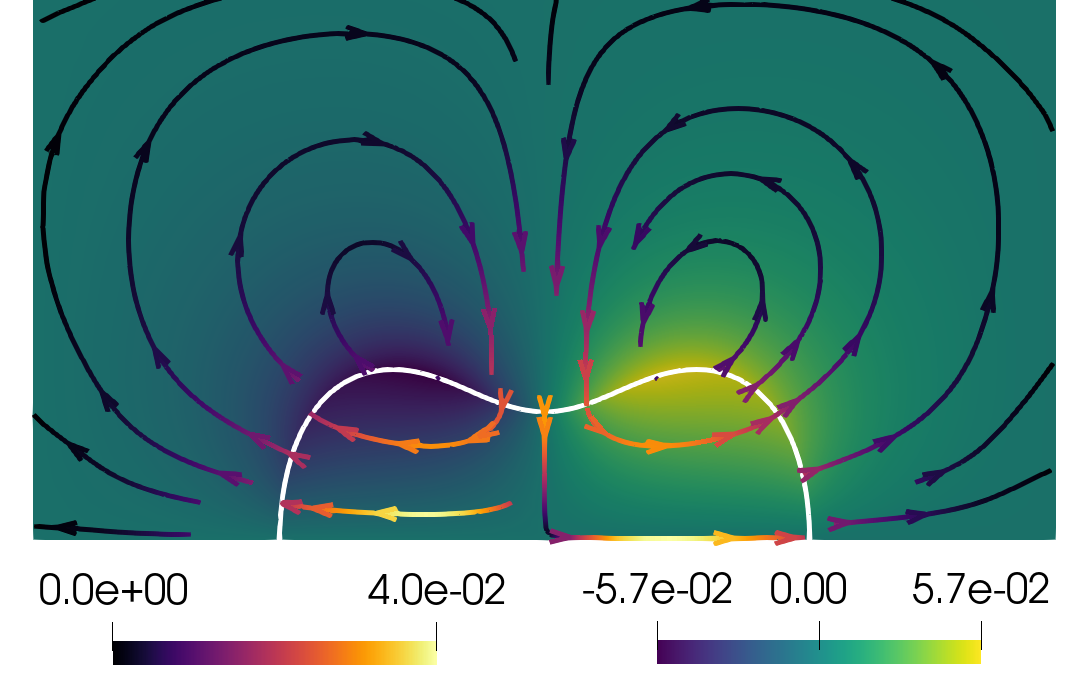}
        \caption{$t=30$}
        \label{fig:example30}
    \end{subfigure}
    \caption{Simulated time series of a chiral force field (inset top left) acting on an initially spherical  viscoelastic surface shown in the $z$-$r$ plane. The streamlines represent the velocity in the plane (left color legend). The background colour shows the velocity in the azimuthal direction $v_\phi$ (right color legend). The chiral force field results in tangential flows  orthogonal to the chiral flows. In this way, a neck forms. Parameters were chosen as $\hat{\tau}_A = \hat{\tau}_S = 100$ and $\hat{\eta}_A = \hat{\eta}_S = 10$.}
    \label{fig:example}
\end{figure}

In the analysis of our simulations, we mainly study two quantities:
\begin{enumerate}
    \item The dimensionless radius $r$ of the surface at the equator, i.e. at $z=0$. This quantity is used to measure the ability of the chiral force field to induce a constricted neck region, indicated by values below one. 
    \item The surface concentration $c$, which models the coarse-grained density of a passive surface bound molecular species. We are especially interested in the  deviation of the equatorial concentration from the concentration in the rest of the surface. Therefore, we calculated $\left(c(z=0)-\int_\Gamma c d\Gamma/|\Gamma |\right)$ for each simulation.
\end{enumerate}
An example of a simulation of a viscoelastic surface deforming under the influence of a chiral force field  is given in Fig. \ref{fig:example}. There, the chiral force field induces a chiral flow pattern, so the azimuthal velocity $v_\phi$ switches sign at the equator. The mechanical properties of the surface transforms this chiral flow into orthogonal flows shown in Fig.~\ref{fig:example}. These flows cause the formation  of a neck at the equator. 

\begin{figure}
    \centering
    \begin{subfigure}{\textwidth}
        \centering
        \begin{tabular}{cc}
             \includegraphics[width=\figWidthProfiles]{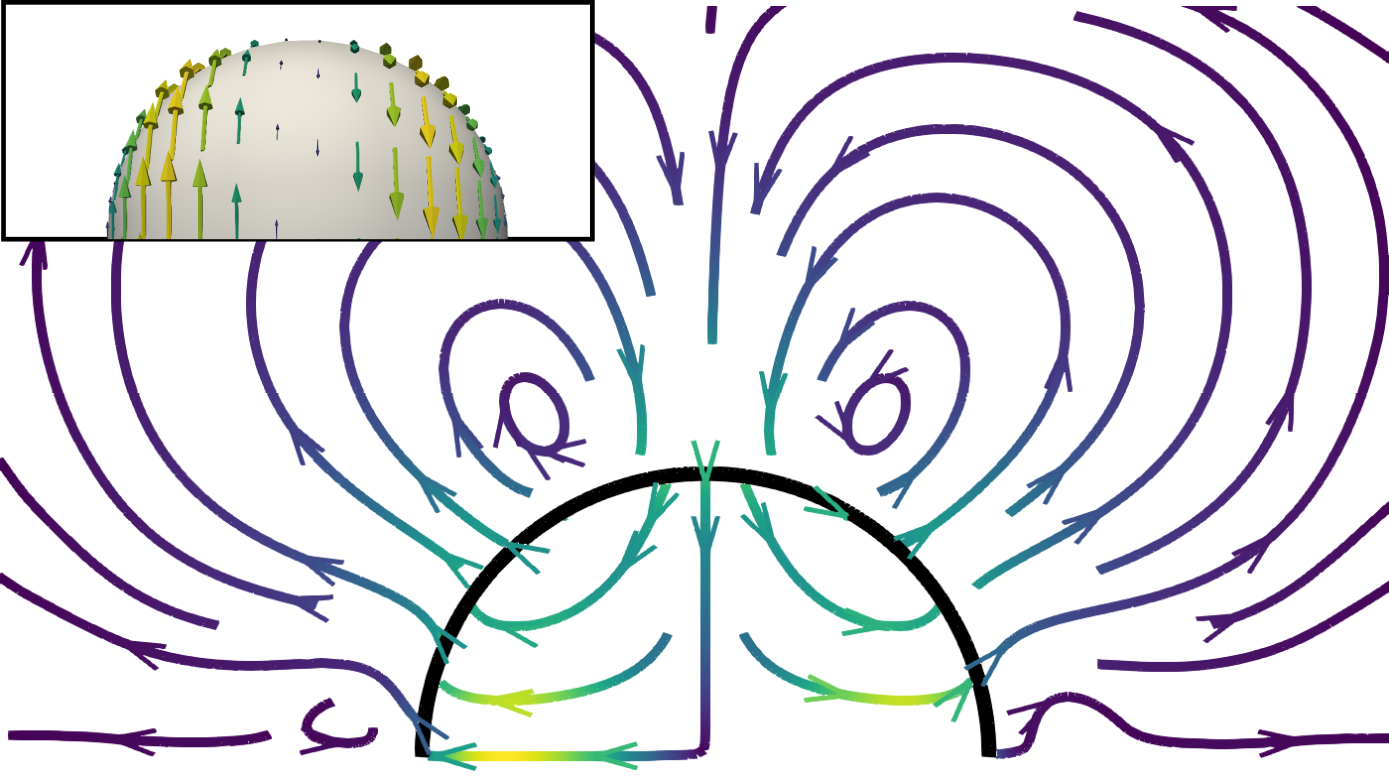}& 
            \includegraphics[width=\figWidthProfiles]{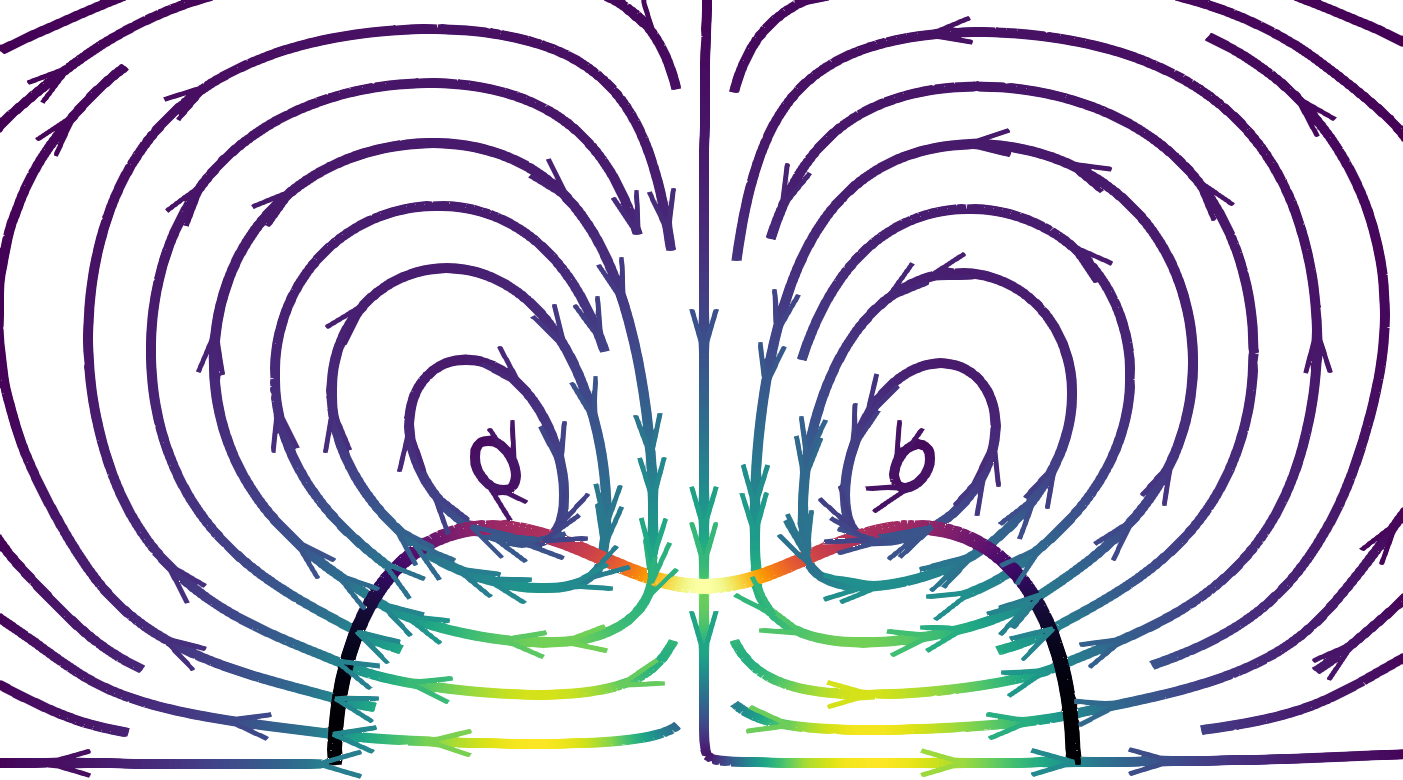} \\
        $t=1$  &  $t=30$ \\
        \end{tabular}
        \captionsetup{width=.9\textwidth}
        \vspace{-0.5cm}
        \caption{Profile A.}
        \vspace{0.5cm}
        \label{fig:profileA}
    \end{subfigure}
    \begin{subfigure}{\textwidth}
        \centering
        \begin{tabular}{cc}
             \includegraphics[width=\figWidthProfiles]{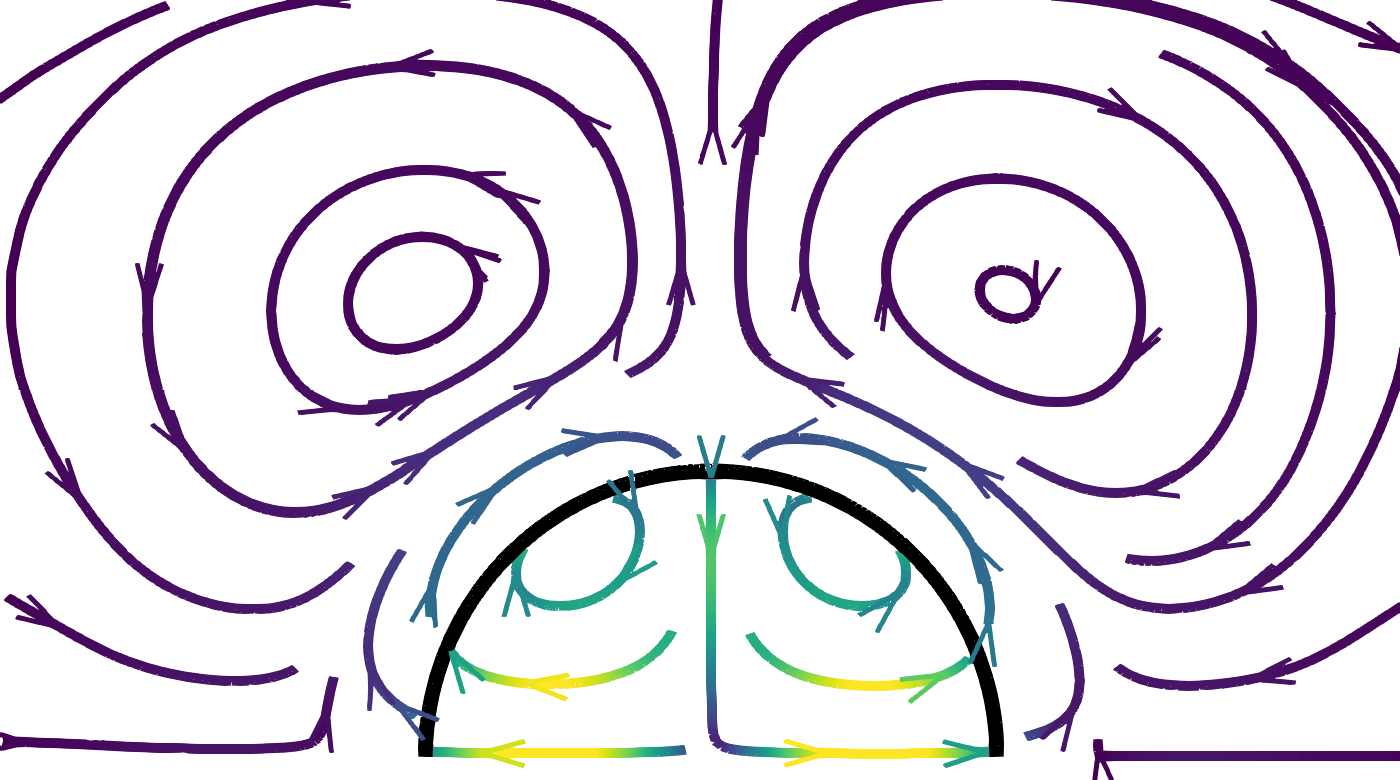}& 
            \includegraphics[width=\figWidthProfiles]{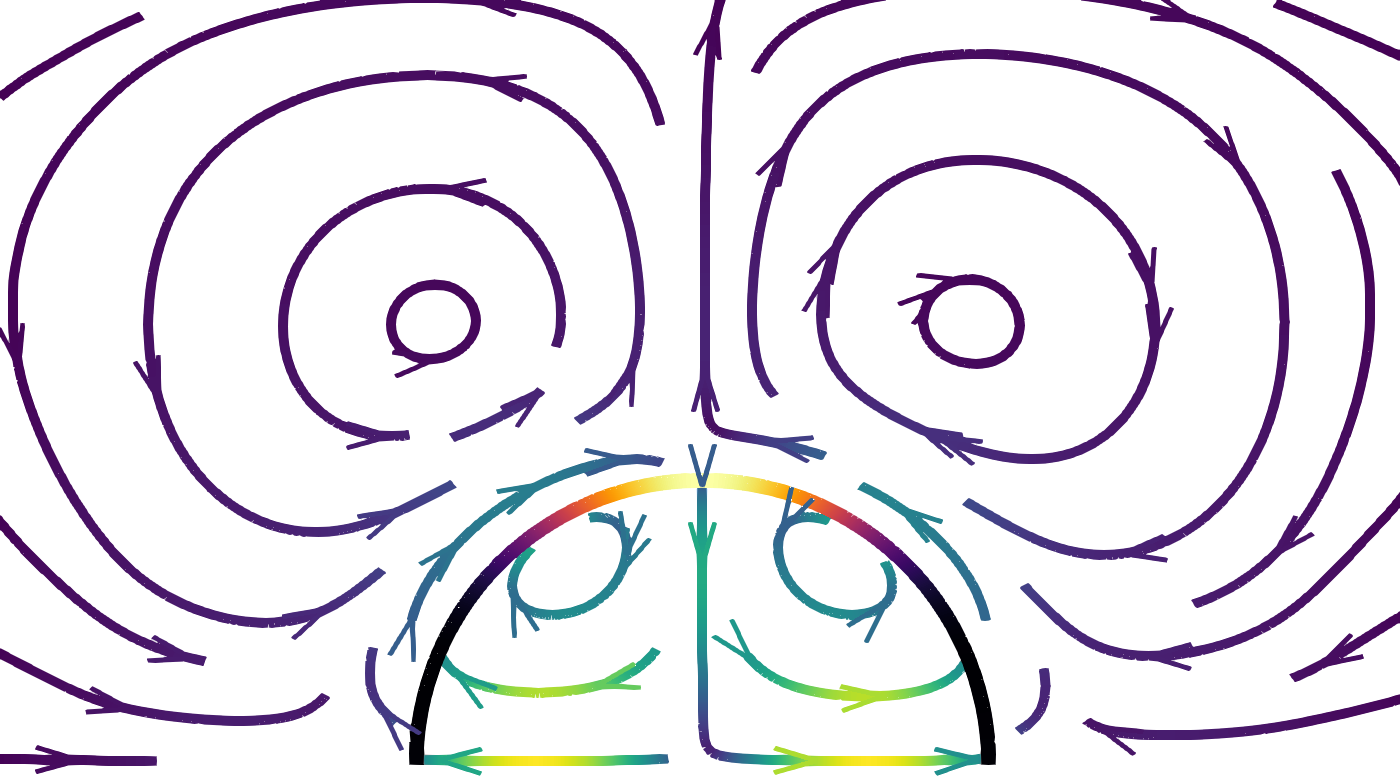} \\
        $t=1$  &  $t=30$ \\
        \end{tabular}
        \captionsetup{width=.9\textwidth}
        \vspace{-0.5cm}
        \caption{Profile B}
        \vspace{0.5cm}
        \label{fig:profileB}
    \end{subfigure}
    \begin{subfigure}{\textwidth}
        \centering
        \begin{tabular}{cc}
             \includegraphics[width=\figWidthProfiles]{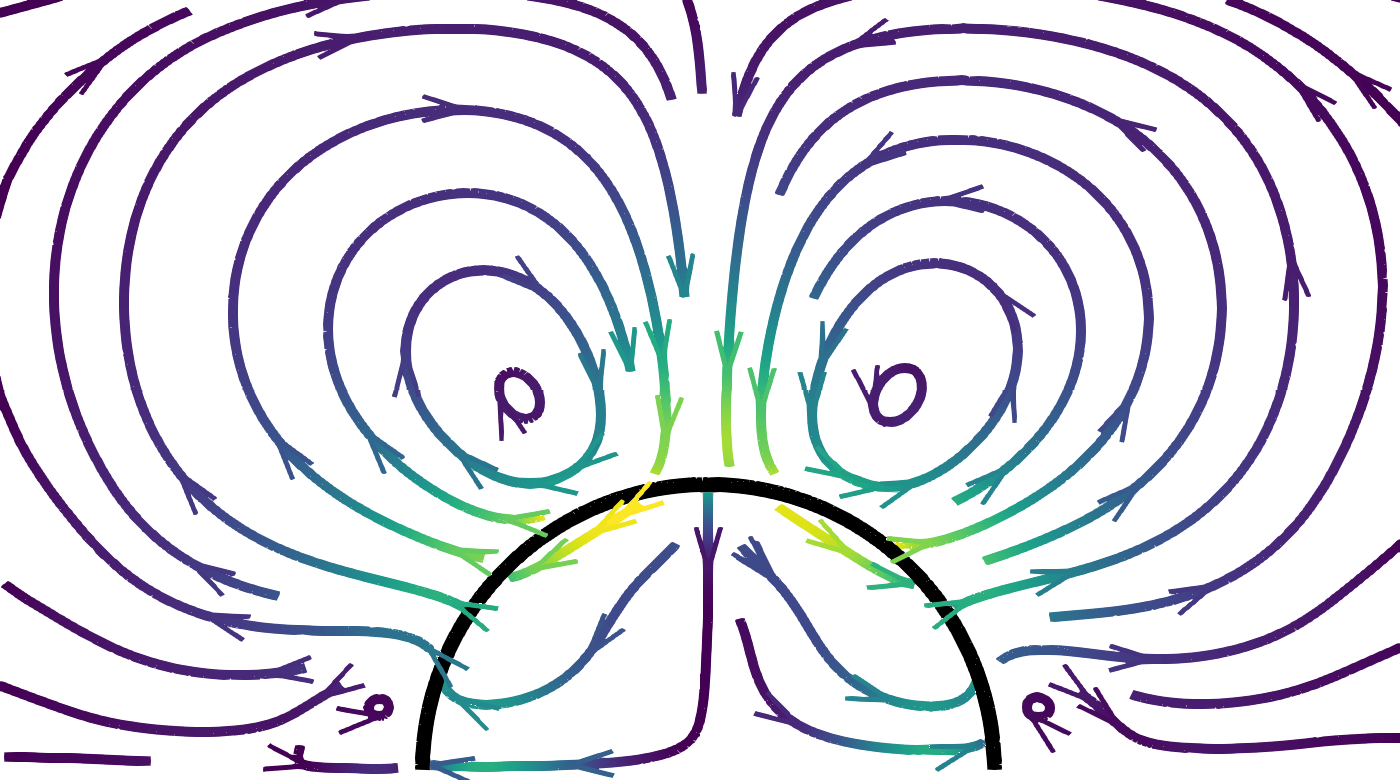}& 
            \includegraphics[width=\figWidthProfiles]{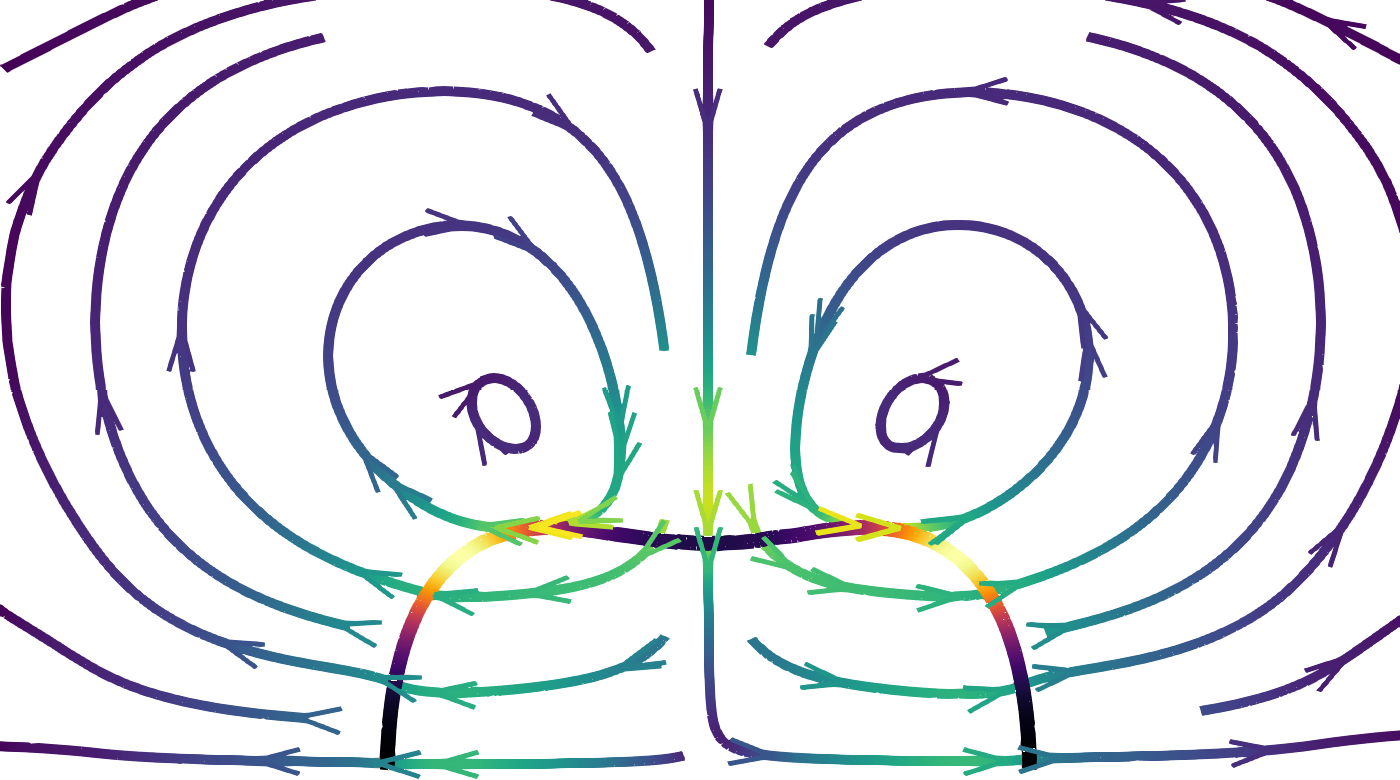} \\
        $t=1$  &  $t=30$ \\
        \end{tabular}
        \captionsetup{width=.9\textwidth}
        \caption{Profile C}
        \label{fig:profileC}
    \end{subfigure}
    \caption{A chiral force field (inset top left) induces three different categories of orthogonal flow profiles in dependence of the relaxation times $\hat{\tau}_A$, $\hat{\tau}_S$. Categories are labelled as  profiles A to C. The streamlines represent the fluid flow, yellow for fast, dark blue for slow. The colour of the surface represents the concentration $c$ of a transported surface quantity, yellow for high, black for low. \textbf{Top:} Profile A, defined by the fact that there are no vortices in the cell and that the maximal velocity is at $r=0$. At later times it can result in the formation of a neck and a ring of high concentration. Parameters: $\hat{\tau}_A = 100$, $\hat{\tau}_S = 100$. \textbf{Middle:} Profile B, defined by the presence of two vortices both inside and outside the cell. At later times a ring of high concentration forms. Parameters: $\hat{\tau}_A = 100$, $\hat{\tau}_S = 0.1$. \textbf{Bottom:} Profile C, defined by having no vortices in the cell and its maximal velocity along the surface. At later times a neck can be formed, but it will not give a ring of high concentration in the centre. Parameters: $\hat{\tau}_A = 1$, $\hat{\tau}_S = 10$. The elastic moduli are in all simulations $\hat{G}_A = \hat{G}_S = 0.1$. }
    \label{fig:profiles}
\end{figure}

\noindent In our simulation, we discovered that the solutions for the fluid velocity could be divided into three categories, which we refer to as flow profiles A to C. These solutions were not only different in appearance, but also resulted in different deformations of the surface and different distributions of the surface concentration $c$, see Fig. \ref{fig:profiles}. Flow profiles are characterised as follows in the different categories: 
\begin{enumerate}[A]
    \item The flow profile has no vortices within the surface and its highest value is along the symmetry axis. The dynamics leads to only mild changes in surface concentration at the centre (Fig. \ref{fig:profileA}).
    \item The velocity profile has two vortices (vortex rings in 3D, respectively) inside and two vortices outside of the surface. Due to the parallel flows towards the equator this profile resulted in the strongest growth of the ring of high concentration $c$. It is therefore the most beneficial profile in increasing the surface concentration at the equator. This profile was only observed in simulations where $\hat{\tau}_A >\hat{\tau}_S$ (Fig. \ref{fig:profileB}).
    \item The flow profile lacks vortices inside of the surface and its highest velocity is between the poles and the equator of the surface. The corresponding dynamics increases the concentration between the poles and the equator. Correspondingly, this pattern is fundamentally different  from a concentration-enriched contractile ring pattern  as observed during cell division \cite{Bonati2022, Reymann2016, Spira2017}. 
    This profile was only observed in simulations where $\hat{\tau}_A < \hat{\tau}_S$ (Fig.~\ref{fig:profileC}).
\end{enumerate}
For the simulations, we choose the diffusion coefficient in Eq.~\eqref{eq:concentrationScaled} to be negligibly small ( $\hat{D}_c = 10^{-4}$). So the dynamics for $c$ is almost exclusively defined by the advective flux, i.e. by the term $\nabla_\Gamma \cdot \vv{v}$. If the surface contracts locally (i.e. $\nabla_\Gamma \cdot \vv{v} < 0$) then $c$ will increase and vice versa. For profiles B and C, the sign of $\nabla_\Gamma \cdot \vv{v}$ is mainly decided by the flows parallel to the surface $\Gamma$. In the case of profile B the surface is compressed at the equator 
and a high concentration spot is formed there. In the case of profile C, the surface is compressed at an angle of approximately $\pm 40^\circ$ with the $r$ axis resulting in two peaks in concentration at these locations. In case of Profile A, we only get a ring of high concentration due to the equatorial compression when a neck is formed. Before that, no strong peaks in concentration are present.


\subsection{Neck formation for equal shear and areal relaxation times}
\label{sec:viscoelastic}
To study the influence of the viscoelasticity on the dynamics, we first assume the areal and shear components to be equal, so $\hat{\tau}_A = \hat{\tau}_S  = \hat{\tau}$ and $\hat{\eta}_A = \hat{\eta}_S = \hat{\eta}$. 
It is noteworthy that for small $\hat{\tau}$ the surface dynamics approaches the viscous limit and for $\hat{\tau}, \hat{\eta} \gg 1$ the surface dynamics approaches the elastic limit of the viscoelastic spectrum. 
Choosing parameters $\hat{\tau}$ in the interval $[10^{-2}, 10^{3}]$ and $\hat{\eta}$ in the interval $[10^{-1}, 10^{3}]$, we exclusively obtain simulations results with flow profile A. 

\begin{figure}
    \centering
    \includegraphics[width=0.49\textwidth]{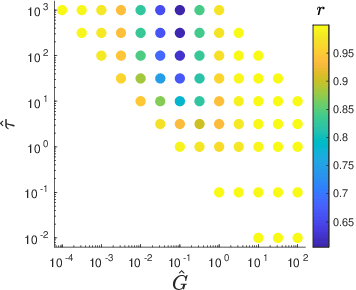}
    \includegraphics[width=0.49\textwidth]{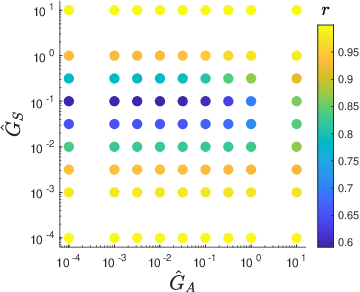}
    \caption{Equatorial radii of viscoelastic surfaces under chiral forces. Each dot represents one simulation. 
    \textbf{Left:} For varying elastic modulus $\hat{G} = \hat{G}_A = \hat{G}_S$ and relaxation time  $\hat{\tau}=\hat{\tau}_A=\hat{\tau}_S$. Results indicate that deformation increases with $\hat{\tau}$ and there is an optimal elastic modulus $\hat{G} \approx 0.1$ where deformation is maximal. 
    \textbf{Right:} For varying elastic moduli at $\hat{\tau}_A = \hat{\tau}_S = 1000$. Results indicate that the shear elastic component mostly determines deformation, with $\hat{G}_S \approx 0.1$ being optimal to achieve large deformation.
    All results shown are at simulation time $t=30$.}
    \label{fig:Gtau_GbGs}
\end{figure}

In Fig. \ref{fig:Gtau_GbGs} left, the radius of the equator is shown for two different simulation times for various $\hat{\tau}$ and $\hat{G} = \hat{\eta}/\hat{\tau}$. 
For either time, we barely see any change in $r$ for small values of $\hat{\tau}$, i.e. for a dominantly viscous regime. We conclude that the viscous component of the surface by itself does not induce any change in shape under chiral forces. Intuitively, this makes sense as the viscosity only acts as friction. But it can also be shown analytically, as is done in appendix \ref{app:viscousForce}. Here we find that the resulting viscous force will be in the opposite direction of the chiral forces and contains no normal components. 

Furthermore, the smallest radii at the equator, $r$, are found along a line of constant elastic modulus $\hat{G}$. So there is an optimal elastic modulus which leads to the highest deformation. 
The highest deformations along this line are observed for large relaxation times $\hat{\tau}$. From this we can conclude that a purely elastic surface (i.e. large $\hat{\tau}$) gives the strongest deformation. The optimal elastic modulus $\hat{G}$ for deformation is higher for earlier simulation times (data not shown) and converges to $\hat{G} = 0.1$, i.e. $\hat{\tau} = 10 \hat{\eta}$ for later simulation times (see Fig. \ref{fig:Gtau_GbGs} left). Our conjecture is, that this phenomenon emerges because  two antagonistic effects are counteracting each other; while a high elasticity causes the surface to deform faster initially, it stops the deformation earlier as the higher shear stresses counteract the chiral forces and reduce the chiral velocity $v_\phi$. Therefore, an intermediate elastic modulus $\hat{G}$ presents the optimal choice for strong deformations.

\subsection{Elasticity-dominated surface dynamics in dependence of shear and areal elastic moduli $\hat{G}_A, \hat{G}_S$}
\label{sec:viscosities}
In Sec. \ref{sec:viscoelastic}, we concluded that the deformation is caused by the elastic component. So to study the influence of the shear and areal elasticity, we ran simulations with a dominant elastic element, i.e. for parameter regimes with large elastic relaxation times fixed at $\hat\tau_{A/S}= 10^3$ and independently varying elastic moduli $\hat{G}_A, \hat{G}_S \in [10^{-4}, 10]$. So the surface viscosities $\hat{\eta}_A$ and $\hat{\eta}_S$ are in the interval $[0.1, 10^4]$. 

The radius $r$ at the equator of the numerical solutions at time $t=30$ is given in Fig. \ref{fig:Gtau_GbGs} right. We find that there is an optimal value for the shear elastic modulus for which the radius $r$ decreases the most. The optimum is $\hat{G}_S= \hat{\eta}_S/\hat{\tau}_S = 0.1$, which is the same optimum found for the elastic modulus $\hat{G}$ in Sec. \ref{sec:viscoelastic}. We infer that the found optimum in Sec. \ref{sec:viscoelastic} was not an optimum for both elastic moduli, but only for the shear elastic modulus. We also observe that increasing the areal elastic modulus $\hat{G}_A$ slightly increases the radius $r$, so the areal elasticity seems to counteract the formation of a neck. To explain this, consider that the initial shape is a sphere, so any deformation at constant volume will increase the surface area, which is resisted by the areal elasticity. 



\subsection{Surface dynamics for distinct shear and areal relaxation times}
\label{sec:relaxationTimes}
\begin{figure}
    \centering
    \includegraphics[width=0.8\columnwidth]{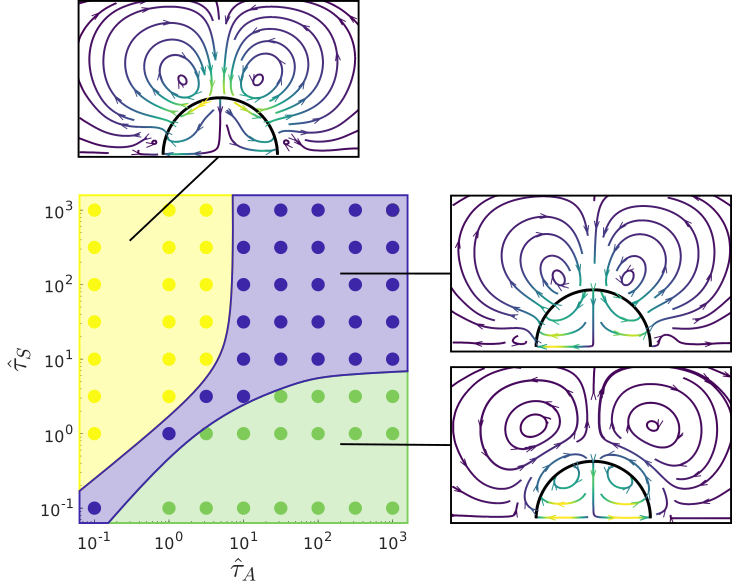}
    \includegraphics[width=0.49\columnwidth]{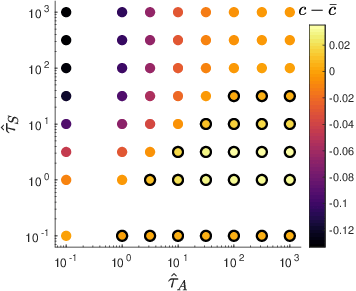}
    \includegraphics[width=0.49\textwidth]{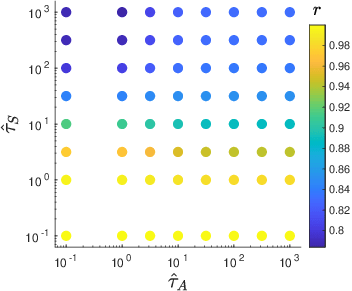}
    \caption{Study of various relaxation times $\hat{\tau}_A, \hat{\tau}_S$ with constant elastic moduli $\hat{G}_A = \hat{G}_S = 0.1$. Each dot represents a numerical solution. \textbf{Top:} Flow profiles A (dark blue), B (green) and C (yellow) at $t=10$. For each profile the example from the left column of Fig. \ref{fig:profiles} is included. \textbf{Bottom left:} The equatorial concentration minus the average concentration $\bar{c}$ at $t=10$. The dots with a black outline indicate when $c-\bar{c}>0$, which only occurs if $\hat{\tau}_A > \hat{\tau}_S$. The strongest increase in equatorial concentration is found for simulations that displayed flow profile B. \textbf{Bottom right:} Equatorial radius at $t=20$. Larger $\hat{\tau}_S$ result in a smaller radius $r$.}
    \label{fig:tauBtauS}
\end{figure}

In the following, we vary the relaxation times $\hat{\tau}_A$ and $\hat{\tau}_S$ independently of each other with fixed elastic moduli $\hat{G}$. We choose the elastic moduli to be equal to the found optimum in Sec. \ref{sec:viscoelastic}, so $\hat{G}_A =\hat{G}_S = \hat{G}= 0.1$. 
With these parameter choices, we observe all three flow profiles (Fig. \ref{fig:tauBtauS} top). When the relaxation times do not differ too much, the flows display profile A. However when the relaxation times are different, we find diverse profiles. 
For a large $\hat{\tau}_A$ and small $\hat{\tau}_S$ the flows display profile B (Fig.~\ref{fig:profileB}). Characteristically, the tangential flows transport the surface bound species $c$ towards the equator resulting in a  higher concentration at the equator over time, see Fig. \ref{fig:tauBtauS} left. This behaviour is qualitatively similar to the enrichment of actin at the cell equator during cell division \cite{Reymann2016}. 

For a large $\hat{\tau}_S$ and small $\hat{\tau}_A$, the flows display profile C (Fig.~\ref{fig:profileC}). This flow profile transports the surface quantity  away from the equator and forms two high concentration spots between the equator and the poles. Correspondingly, this flow pattern  counteracts the formation of a high concentration equatorial ring. However the flows induced by the chiral forces do result in the largest decrease in $r$ at the equator, as is shown in Fig. \ref{fig:tauBtauS} right. This confirms the insight of the previous section that the surface is mainly deformed by the shear elastic component of the stress. Moreover, increasing the areal relaxation time increases $r$. So this supports the result in Sec. \ref{sec:viscosities} that the areal elasticity counteracts the formation of a neck.

To study the influence of the geometry, we run the same simulations as before, but now taking an oblate and a prolate as initial surface $\Gamma$. The oblate and prolate have dimensionless radii $1.25$, $1.25$, $1$ and $0.75$, $0.75$, $1$ respectively. The simulations with the oblate and prolate as initial shape showed the exact same phase diagram of the flow profiles at simulation time $t=10$ as the simulations with a sphere as initial shape (Fig. \ref{fig:tauBtauS} Top). For each profile, the corresponding dynamics is  similar to those observed for the simulations with the sphere as initial condition. From this we conclude that a small change in geometry does not affect the dynamics of the system.

\subsection{Chiral forces may stabilise ring structures in active pattern-forming viscoelastic surfaces}
\label{sec:contractileRing}
In previous work, active viscous surfaces were studied as a minimal model of the self-organisation of the cellular actin cortex \cite{Lucas_paper, Bonati2022, Mietke2019a, Mietke2019}. There, active surface tension was assumed to be determined by a surface concentration $c$ of molecular regulators, see Eq.~\eqref{eq:active tension}.
It was shown that the interplay of surface concentration, tension and flows can lead to pattern formation \cite{Mietke2019, bois11, kuma14, salb17, wagn16, salb09, bert14, gros19}. From a biological point of view, the most notable patterning is the formation of a region of high concentration around the equator of the cell, i.e. a ring mode, which might in turn lead to the formation of a neck. This pattern resembles the formation of a contractile actin-cytoskeletal ring in a dividing animal cell \cite{reym16, Spira2017}. 
However, for models of the active cortical surface, it was reported that for high activity (i.e. high $\hat{\xi}$ in Eq.~\eqref{eq:activeSurfTenScaled}) such a contractile ring will be unstable in the sense that it does not remain at the equator, but slips towards one of the poles over time, resulting in a polar mode \cite{Lucas_paper, Bonati2022, Mietke2019} (see Appendix \ref{app:correlation} for the definitions of the polar and ring modes and our method to compare these).

Here, we test the influence of chiral forces on the formation and stability of a concentration-rich ring of cortical regulators and explore the possibility that the ring slipping is prevented by flows similar to flow profile category B. As seen earlier, this flow profile exhibits strong flows towards the equator (Fig.~\ref{fig:profiles}). This can stimulate the build-up of a surface bound species at the equator, which could prevent the slipping of a contractile ring. To study this phenomenon, we choose a more physically motivated chiral force field. The earlier definition in Eq. \eqref{eq:chiralForce} depends on the shape of the initial surface to have a switch in sign at the equator. However, current literature suggests that chiral flows are caused by local torques \cite{Sase1997,Ali2002,Mizuno2011,Naganathan2014}. These local torques come from the actin filaments, which rotate due to their helix structure when myosin motors exert forces on them \cite{Naganathan2014, Naganathan2016, Middelkoop2021}. So a more physical model would assume that the chiral force field is dependent on the concentration of molecular tension regulators (myosin). Furthermore, if the concentration of actin filaments and tension regulators were uniformly distributed, then the torques caused by the actin filaments would balance each other.  In turn, as a first model for a concentration dependent chiral force field, we propose a force that depends on the gradient of $c$ along the surface,
\begin{equation}
    \vv{f}_c = \alpha \frac{R^3}{c_0} \nabla_\Gamma c \cross \vv{n}.
    \label{eq:newChiral}
\end{equation}
The fraction $\frac{R^2}{c_0}$ is needed to keep the unit of $\alpha$ the same as in the previous definition of the chiral force in Eq.~\eqref{eq:chiralForce}. 
If there is a ring of high concentration around the cell, then we obtain a similar chiral counter-rotating field as before (see Eq.~\eqref{eq:chiralForce}). Even though we use a different definition for the chiral force as in Secs. \ref{sec:viscoelastic}-\ref{sec:viscosities}, we will still use those results as guidance. 

In our simulations, we choose $\hat{\tau}_S = 5$ and $\hat{\tau}_A = 5\cdot 10^3$. For the chiral force defined in \eqref{eq:chiralForce}, these parameters resulted in numerical solutions with profile B, as is shown in Sec. \ref{sec:relaxationTimes}.
Consistent with estimated parameters ranges found in biological cells (see Appendix \ref{app:biology}), we anticipate the viscosity of the cell cortex to dominate the cytoplasmic viscosity and choose $\hat{\eta}_A = \hat{\eta}_S = 10$, and a negligible viscosity of the surrounding medium, $\frac{\eta_0}{\eta_1}=0.1$. 
Using \cite{Mietke2019} as a guidance, we choose the scaled activity $\hat{\xi}=0.02$ and scaled diffusion coefficient $\hat{D}_c = 2\cdot 10^{-4}$ such that the steady state is unstable and we expect a polar mode for the concentration when there is no chiral force field.
As we want to study ring slipping, we choose a ring mode as initial condition. To make the solution less biased, we add a perturbation to it, such that the initial concentration is $c_i(t=0) = 1 + 10^{-4} \left( \frac{1}{2} \left( 3 \cos^2(\theta_i) -1 \right) + \delta_i \right)$, where the index $i$ indicates the $i$th grid point, $\theta_i$ is the angle w.r.t. $z$-axis of the $i$th grid point and $\delta_i$ is a uniform random variable between $-1$ and $1$. 

\begin{figure}
    \centering
    \begin{subfigure}{\textwidth}
    \centering
        \begin{tabular}{ccc}
            \includegraphics[width=0.31\textwidth]{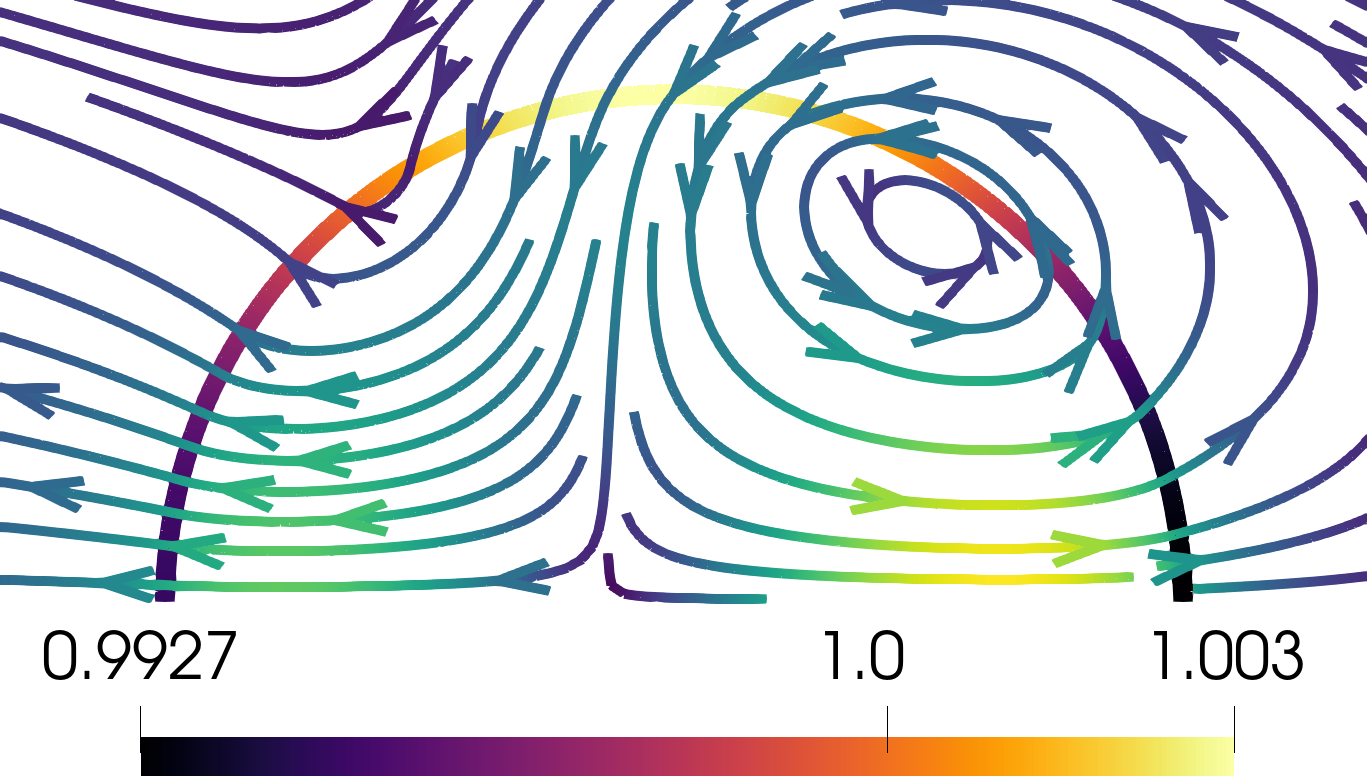} & \includegraphics[width=0.31\textwidth]{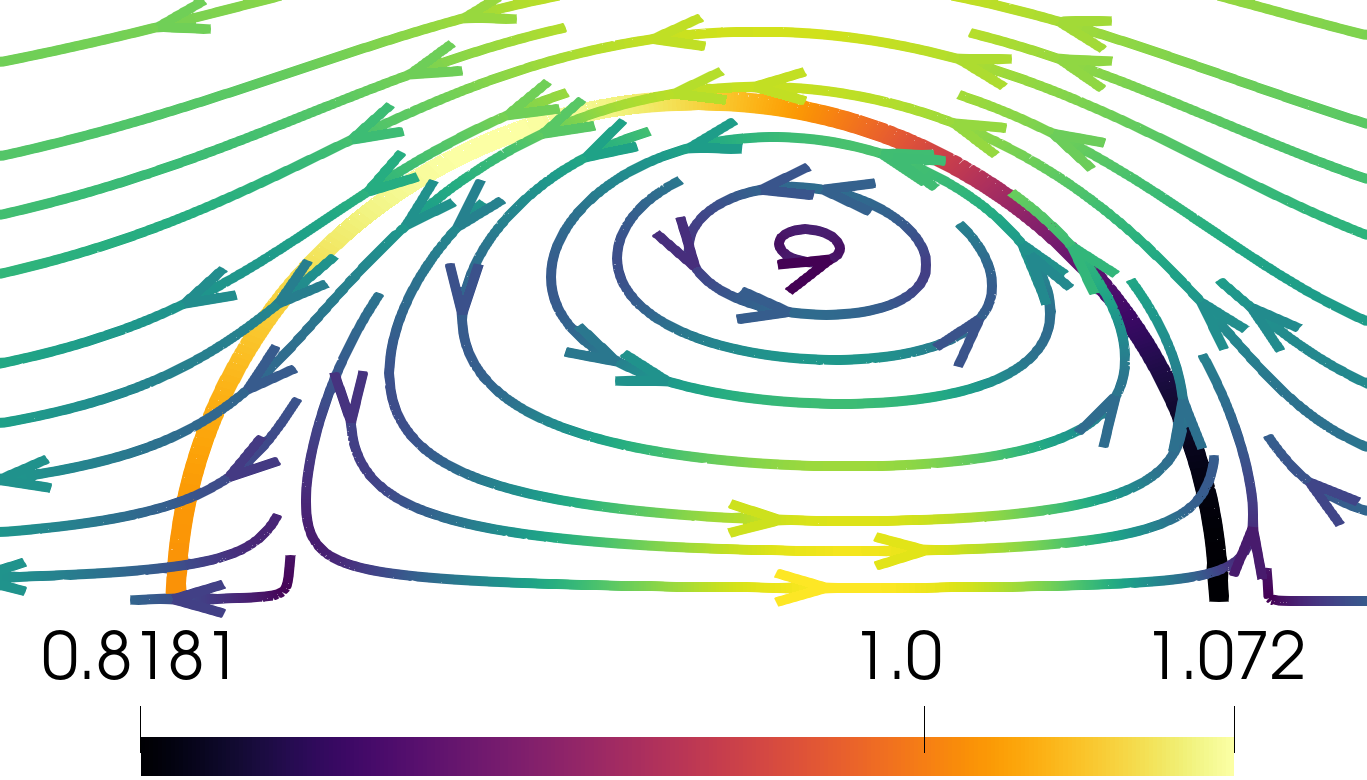} & \includegraphics[width=0.31\textwidth]{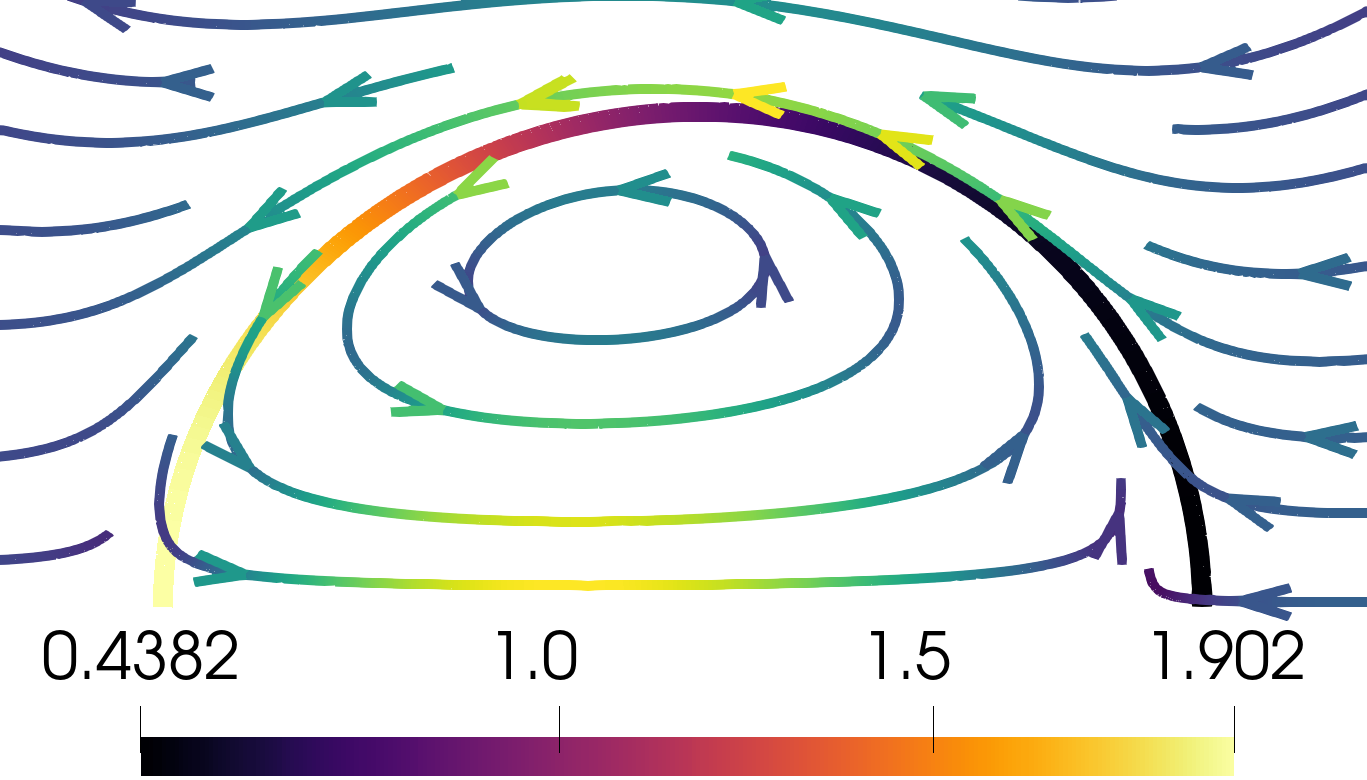} \\
            $t=40$ & $t=130$ & $t=170$
        \end{tabular}
        \caption{Active surface tension.}
        \vspace{0.3cm}
        \label{fig:ringSlipWOchiral}
    \end{subfigure}
    
    \begin{subfigure}{\textwidth}
    \centering
        \begin{tabular}{ccc}
            \includegraphics[width=0.31\textwidth]{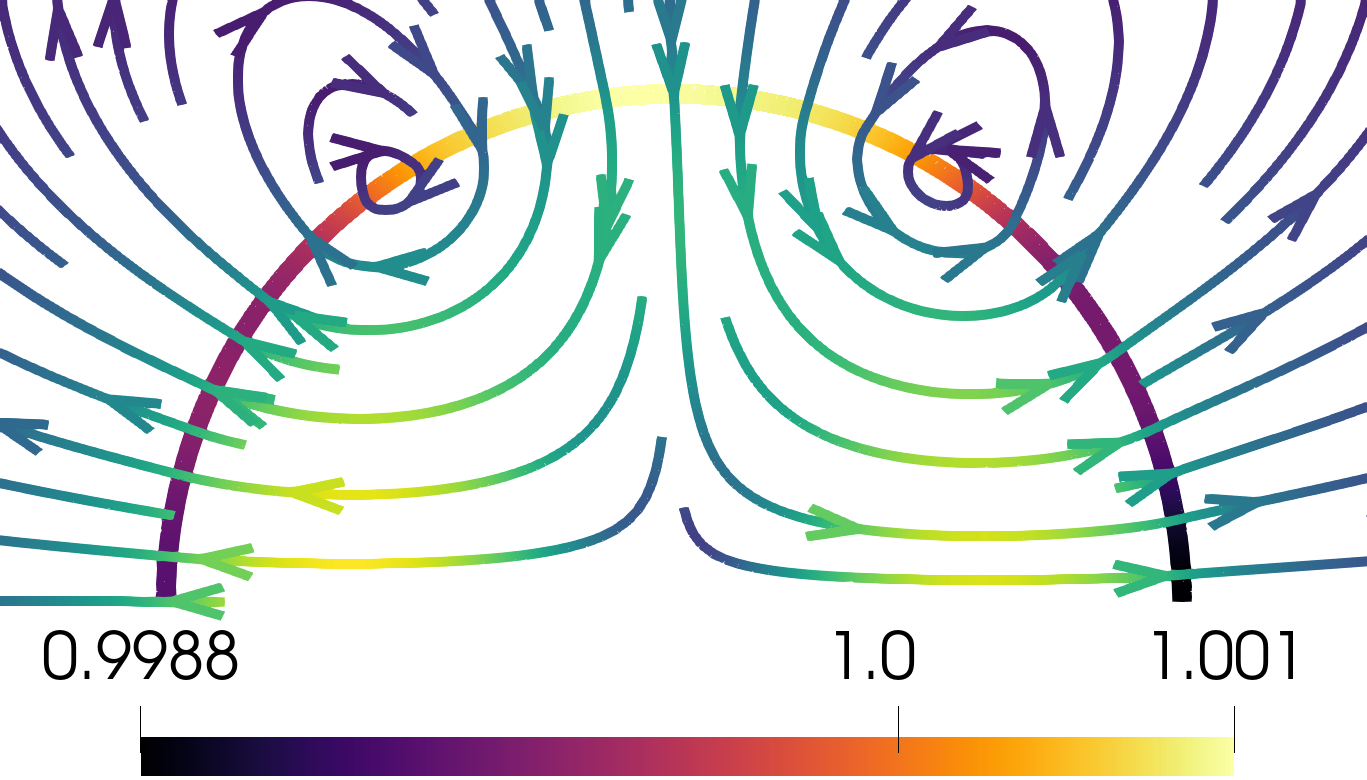} & \includegraphics[width=0.31\textwidth]{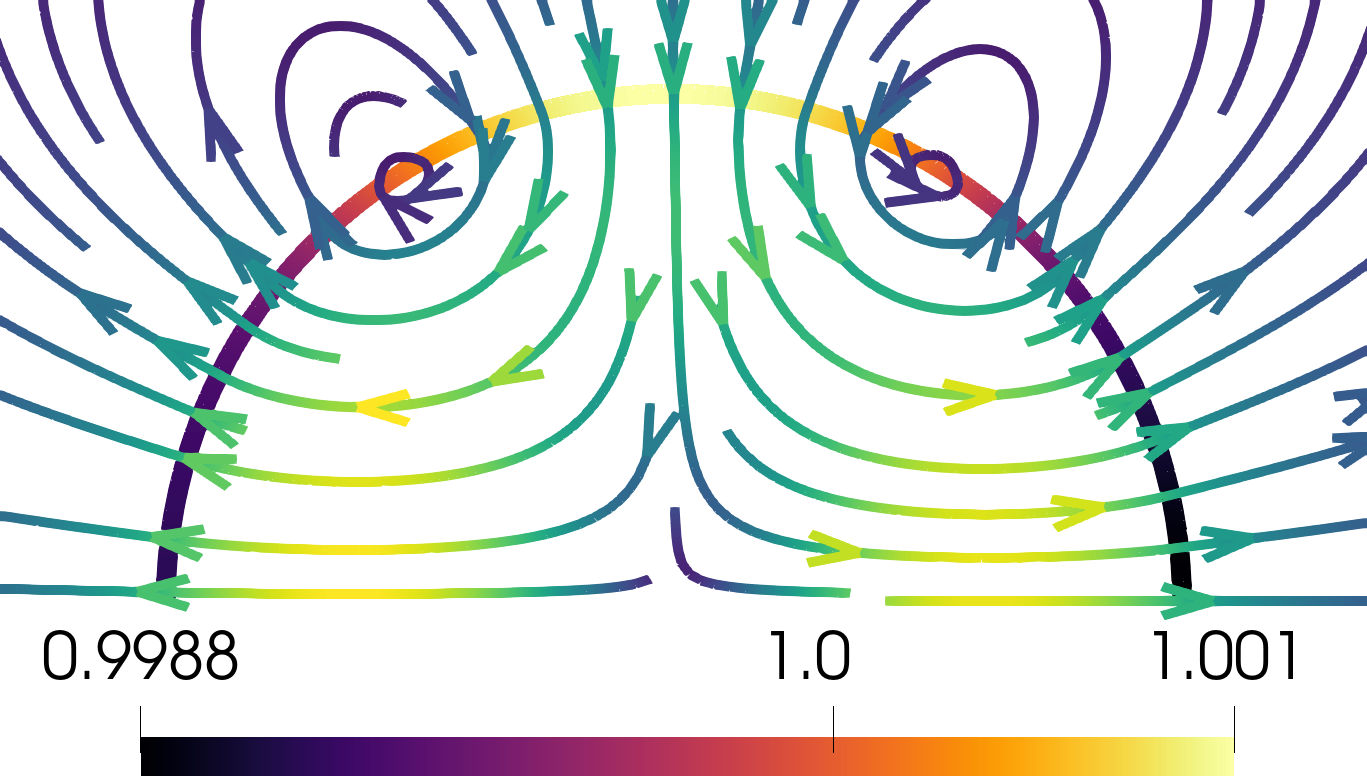} & \includegraphics[width=0.31\textwidth]{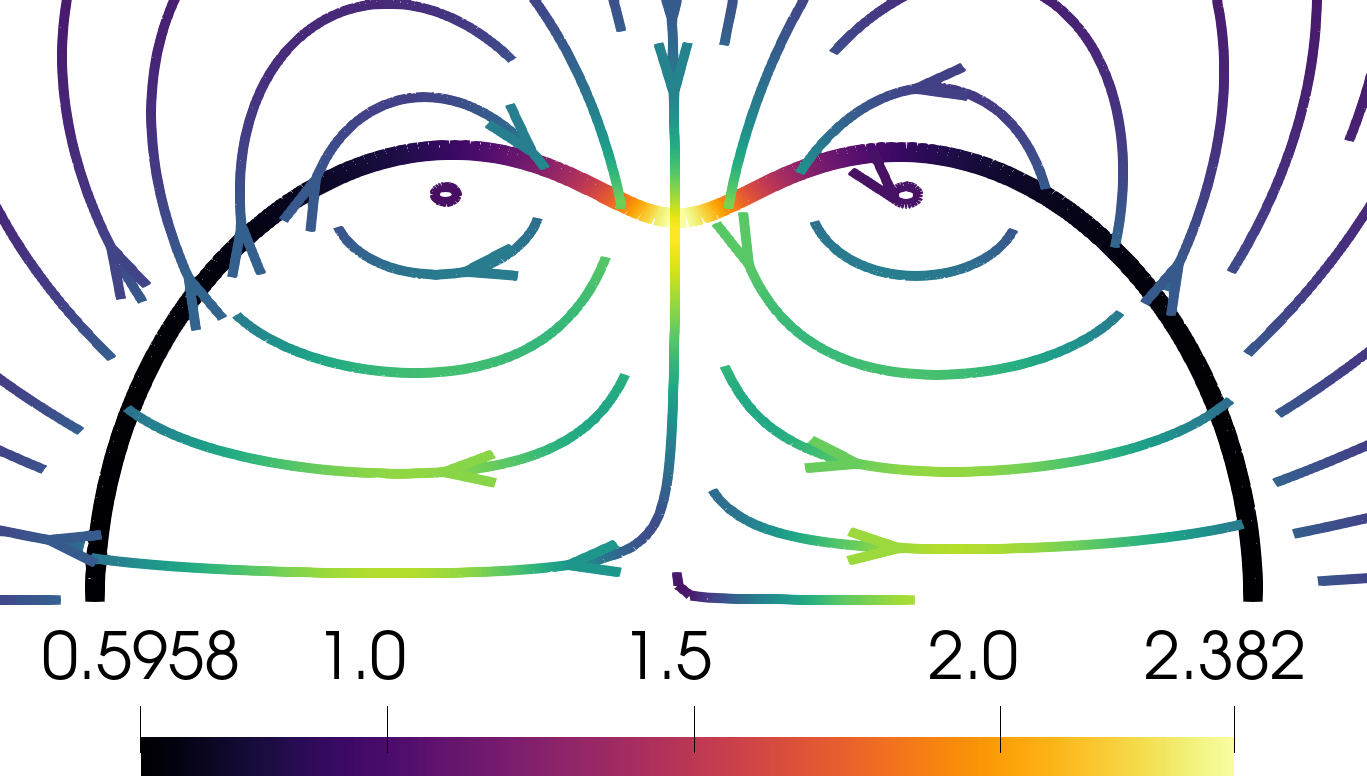} \\
            $t=40$ & $t=130$ & $t=475$
        \end{tabular}
        \caption{Chiral force field.}
        \vspace{0.3cm}
        \label{fig:ringSlipChiral}
    \end{subfigure}
    
    \begin{subfigure}{\textwidth}
    \centering
        \begin{tabular}{ccc}
            \includegraphics[width=0.31\textwidth]{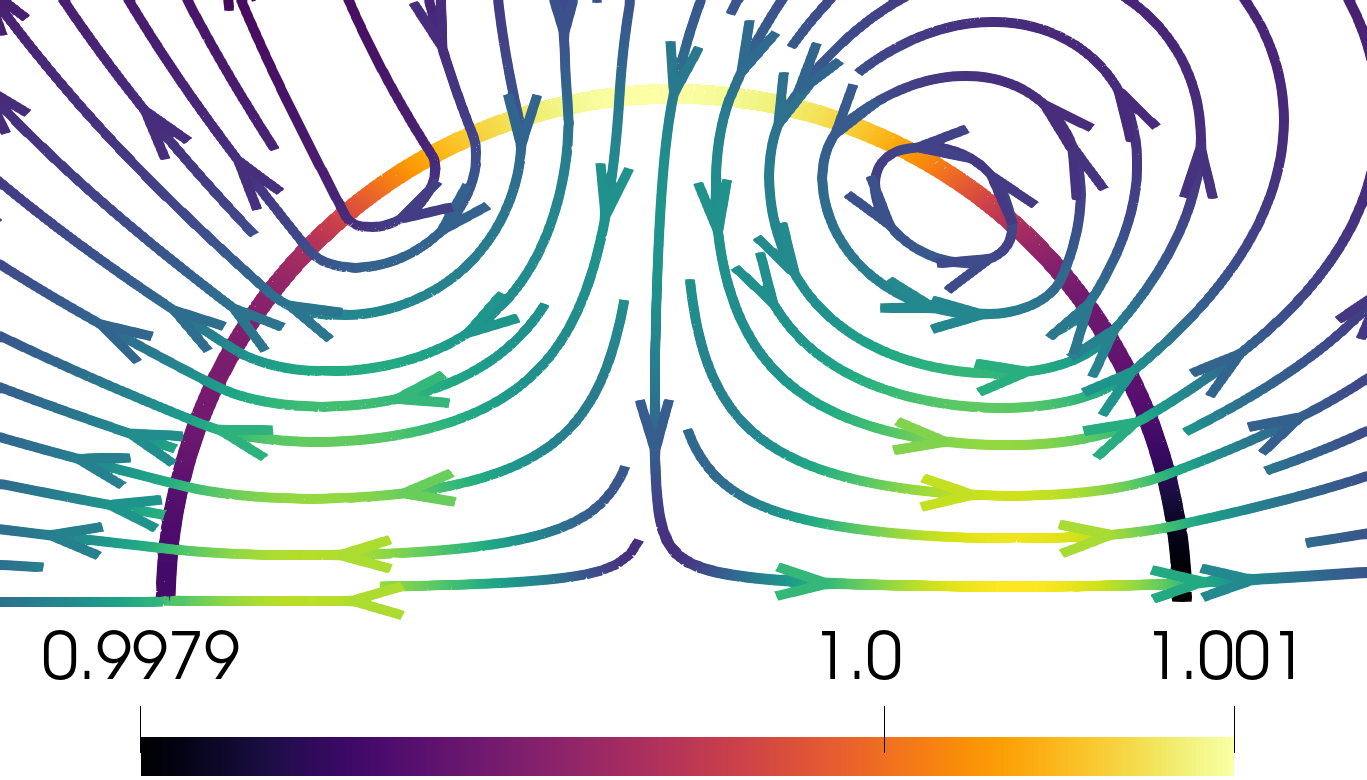} & \includegraphics[width=0.31\textwidth]{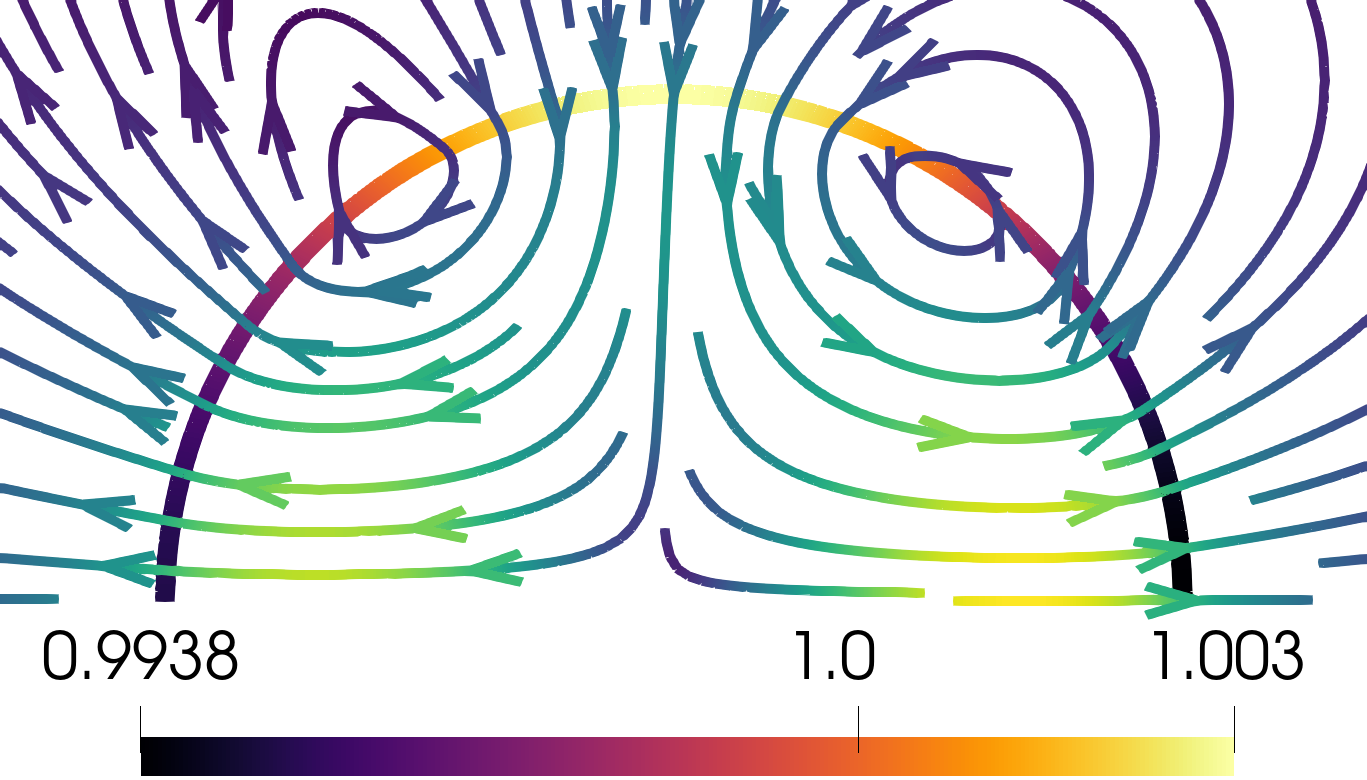} & \includegraphics[width=0.31\textwidth]{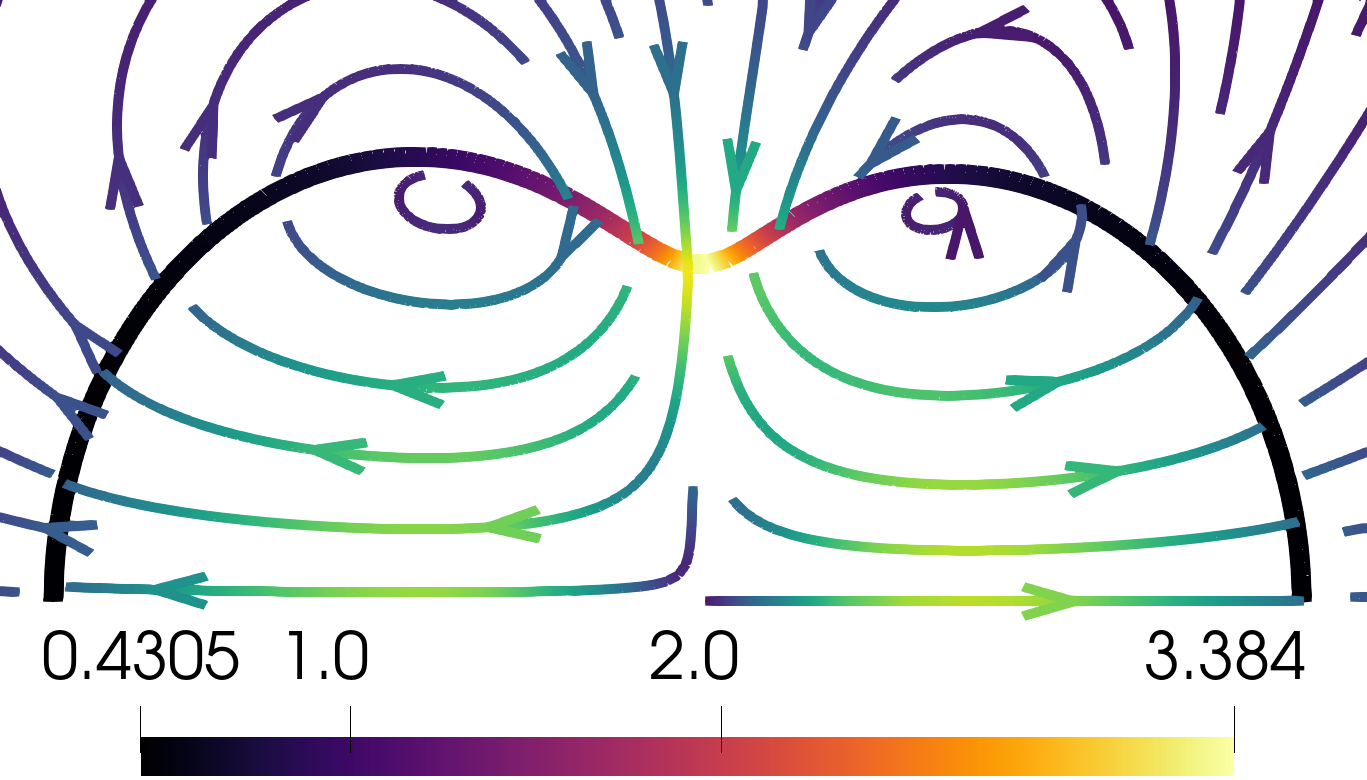} \\
            $t=40$ & $t=130$ & $t=252$
        \end{tabular}
        \caption{Active surface tension and chiral force field.}  
        \label{fig:ringSlipActiveChiral}
    \end{subfigure}
    \includegraphics[width=0.49\textwidth]{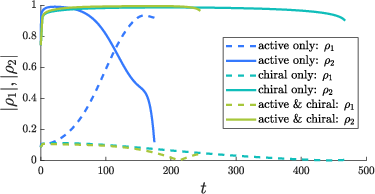}
    \includegraphics[width=0.49\textwidth]{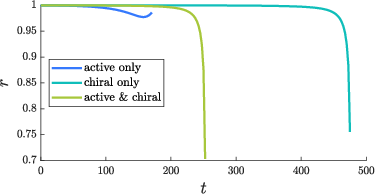}

    \caption{Active pattern formation on viscoelastic surfaces in three different conditions. \textbf{(a)} Having only active surface tension leads to formation of a volatile high-concentration ring which  slips to the left pole. \textbf{(b)} Having only chiral force field leads to a stable high-concentration ring and neck formation. \textbf{(c)} The combination of active surface tension and chiral force field strengthens the ring formation and leads to stronger and earlier neck formation. 
    The colour on the surface represents the concentration $c$. The streamlines represent the velocity with respect to the average velocity of the cell, yellow for high, blue for low. 
    \textbf{Bottom left:} The absolute value of the correlation coefficients (Appendix \ref{app:correlation}) of the polar mode $\rho_1$ and ring mode $\rho_2$. The time domains are chosen such that the surface has not deformed too much from a sphere (Appendix \ref{app:correlation}). \textbf{Bottom right:} Equatorial radius for all three simulations. The time domains are chosen the same as those in Figs. \ref{fig:ringSlipWOchiral}-\ref{fig:ringSlipActiveChiral}.  If there is no chiral force field then the high concentration ring slips ($|\rho_1|\rightarrow 1$, $|\rho_2|\rightarrow 0$). Otherwise the ring is stabilised and a neck forms ($r(z=0)$ decreases).
    Parameters: $\hat{\tau}_A = 5000$, $\hat{\tau}_S = 5$, $\hat{\eta}_A = \hat{\eta}_S = 10$, $\hat{D}_c = 2\cdot 10^{-4}$, $\frac{\eta_0}{\eta_1} = 0.1$.}
    \label{fig:ringSlip}
\end{figure}

We run three simulations, one with only the chiral force field, one with only active surface tension  and one with both. 
In the case of only active tension (Fig.~\ref{fig:ringSlipWOchiral}, Movie 1 in SI) an equatorial ring builds up, but is not stable and slips to one of the poles over time. 
In the second simulation, we use a concentration-dependent chiral force without tension  (Fig.~\ref{fig:ringSlipChiral}, Movie 2 in SI). 
As was reported in Sec. \ref{sec:relaxationTimes}, the chiral force results in flows towards the equator. These flows transport the surface bound species towards the equator, increasing the concentration even further. 
This process, in combination with the concentration-dependent chiral force (see Eq.~\ref{eq:newChiral}) comprises a positive feedback loop.  Accordingly, we observe the self-organisation of the chiral forces and surface concentration leading to a stable ring pattern and a neck formation at late times. 
The results illustrate that chiral force-feedback provides a new mode of mechano-chemical pattern formation which does not require active surface tension.

In the last simulation, we combine active tension and chiral forces (Fig.~\ref{fig:ringSlipActiveChiral}, Movie 3 in SI). 
Again, a ring pattern is formed. But this time it is even enhanced by the concentration-dependent surface tension leading to a faster built-up of concentration and earlier neck formation. 
In general, for simulations with both active tension and chiral flows we find that the formation of a ring of high concentration is faster for larger $\hat{\xi}$. However if $\hat{\xi}$ becomes too large, then the ring is not stabilised anymore but slips towards one of the poles resulting in a polar mode. 
From this, we conclude that the chiral force field can not only stabilise the contractile ring, but in addition, the chiral forces and active surface tension "cooperate" in the sense that they amplify their mutual influence on surface concentration aggregation and constriction.

All simulations results are shown until the time where the pattern and shape dynamics become too strong, i.e. if the concentration peak becomes too concentrated in space or the surface develops deformations with very high local curvature. In this case, the dynamics cannot be reliably resolved by the numerical grid.

\section{Discussion}
\label{sec:discussion}
Here, we present the first numerical model of a viscoelastic surface deforming under the influence of a counter-rotating force field.  Using a frame-invariant Maxwell model for the material properties of the surface, we implement a type of viscoelasticity where elastic in-plane stresses in the surface are dissipated over characteristic time scales. In this model, areal and shear deformations are characterised by two independent sets of viscoelastic parameters. 

Performing simulations at different parameters, we find that the surface barely deforms if both viscoelastic relaxation times are small (Secs. \ref{sec:viscoelastic} and \ref{sec:relaxationTimes}).  From this we conclude that a purely viscous surface does not change shape under a chiral force field. This is  also consistent with an analytical derivation showing that purely viscous surface stresses give rise to only tangential flows on the surface, see Appendix \ref{app:viscousForce}. 
It is notable to point out the counter-intuitive nature of this fact, as viscous behaviour is typically associated with large deformations.

In addition, we find that increasing the shear relaxation time $\hat{\tau}_S$ is optimal for the formation of a neck, indicating that  the formation of a neck is caused by the shear elastic component (Secs. \ref{sec:viscoelastic} and \ref{sec:relaxationTimes}). 
In contrast, increasing the areal relaxation time $\hat{\tau}_A$ or increasing the areal elastic modulus $\hat{G}_A$ decreases the formation of a neck (Figs. \ref{fig:Gtau_GbGs} right and \ref{fig:tauBtauS}). 
We propose that this observation can be explained as follows: the chiral force field initially induces a shear stress, which leads to the formation of a neck. Neck formation, however increases the total surface area of the cell, which is resisted, in turn, by the areal elasticity (which scales with $\hat{G}_A$ and $\hat{\tau}_A$). 
Furthermore, we found an optimal value of the shear stress $\hat{G}_S = 0.1$ for neck formation.
This can be understood by taking into account that shear elastic stress is on the one hand needed to induce a neck, on the other hand, large shear elastic moduli will resist deformation strongly and thereby reduce the rotational flows caused by the chiral force fields resulting in less deformation.
Aside from the building of a neck, we also found that a ring of high concentration at the surface equator requires a high areal relaxation time $\hat{\tau}_A$ and a low shear relaxation time $\hat{\tau}_S$.
In this parameter regime we illustrate that concentration-dependent chiral forces can lead to self organisation and induce a ring pattern. 
This is to our knowledge the first evidence that chiral force-feedback provides a mode of mechano-chemical pattern formation which does not require active surface tension.

Finally, we added an active surface tension to the model, which has been shown to lead to ring pattern formation on viscous and viscoelastic surfaces which was, however, reported to be transient in the experimentally relevant parameter regime of low cytoplasmic viscosity \cite{Mietke2019a, MietkeThesis2018}. In animal cells, by contrast, persistent ring patterns are observed during cell division \cite{Salbreux2012}. In our study, we show that for large areal viscoelastic relaxation time scales $\hat{\tau}_A$ a ring of high concentration at the surface equator can be stabilised by the flows induced by a chiral force field. Accordingly, the slipping of a pre-formed contractile ring as observed in \cite{Lucas_paper,Bonati2022} is prevented, and a robust neck is formed. Moreover, we report that chiral forces and active tension collaborate on pattern and neck formation in the sense that they amplify their mutual influence on surface concentration aggregation and constriction.

Our results in Secs. \ref{sec:viscoelastic}- \ref{sec:relaxationTimes} were gained from simulations which were mirrored in the $z=0$ plane. To test whether this symmetry emerges naturally from the dynamics, we ran similar simulations including a slight asymmetric perturbation. We found that the perturbation quickly levelled out, such that the solution went back to being mirror symmetric (data not shown).
Another mechanism we tested is the influence of the initial shape of the surface. We found the same categories of flow profiles as in Fig.~\ref{fig:tauBtauS} which displayed qualitatively the same dynamics as described in Sec.~\ref{sec:Results}. So we conclude that the dynamics of the surface mainly depends on the material properties of the surface and not on small changes in shape.

All in all, our results give a qualitative overview over the effects of counter-rotating flows on viscoelastic surfaces. Thereby, we provide a new perspective on how viscoelastic traits may influence pattern formation and deformation of viscoelastic cell cortices in the presence of chiral flows which have been reported by experimental studies on dividing cells \cite{Danilchik2006, Naganathan2014, Blum2018}. 
Most importantly, we give evidence that in the presence of a sufficiently strong shear elasticity, chiral cortical forces can induce neck formation by two distinct mechanisms: i)~the direct shape dynamics induced by orthogonal flows and ii)~tangential flows which lead to a high concentration ring of motor proteins. Further, pronounced neck formation is resisted by area elasticity of the surface. 
We note that the ranges of parameters used in our simulations overlap with estimated parameters ranges found in biological cells (App. \ref{app:biology}). 

For active surfaces, our simulations put forward, that the self-organised formation of a neck in combination with a concentration-rich equatorial ring can be stabilised and promoted by the presence of chiral forces. Our simulations suggest that this phenomenon requires short shear relaxation times scales but long areal relaxation time scales. While this parameter setting has so far not been experimentally verified in the cell, we speculate that active cell surface area regulation, e.g. through exocytosis and endocytosis, may provide an effective areal elasticity that prevails over long time scales. 

Throughout our study, we have focused on either a prescribed force field or a simple first-order dependence on a surface bound concentration. 
In the cell cortex the chiral forces are generated by a non-trivial tension-torque coupling \cite{DELACRUZ20101852,gore2006dna}, and possibly modifications in the actin helix \cite{ngo2015cofilin,mcgough1997cofilin}. 
In the future, it will be interesting to include a more detailed molecularly motivated description of the chiral force field in the model that enables self-organised ring formation and constriction at the equator.
Finally, we note that chiral flows may also emerge in biological processes which are not axisymmetric. Our proposed numerical model can be directly extended to explore the resulting fully three dimensional shape dynamics. 


\section*{ ACKNOWLEDGEMENTS} 
SA and EFF acknowledge support from the German Research Foundation DFG (grant AL1705/6 and FI 2260/5) from DFG Research Unit FOR-3013. 
EFF was further supported by the Heisenberg program – project number 495224622 (FI 2260/8-1) - and the Deutsche Forschungsgemeinschaft under Germany's Excellence Strategy, EXC-2068-390729961, Cluster of Excellence Physics of Life of TU Dresden. Simulations were performed at the Centre for Information Services and High Performance Computing (ZIH) at TU Dresden. We also want to thank Lucas D. Wittwer for his help with the implementation in C++.

\newpage
\appendix
\section{Numerical implementation}
\label{app:numerical implementation}
To solve the system of PDEs in Eqs. \eqref{eq:devSscaled} - \eqref{eq:activeSurfTenScaled} we use an IMEX method for the time derivatives and finite elements for the spatial derivatives. We have two 2-dimensional grids for the Stokes Eqs. \eqref{eq:stokes1scaled}, \eqref{eq:stokes2scaled} and one 1-dimensional curved grid for the surface Eqs. \eqref{eq:devSscaled}, \eqref{eq:trSscaled} and \eqref{eq:concentrationScaled}. To couple the surface and bulk equations we use the Arbitrary Lagrangian-Eulerian (ALE) approach, as is described in \cite{mokbel2020ale} and \cite{DeKinkelder2021}. 

\subsection{Time integration}
The complete system of equations is split into several sub-problems that are solved subsequently in each time step: Stokes equations (Eqs. \eqref{eq:stokes1scaled}, \eqref{eq:stokes2scaled}), the concentration equation \eqref{eq:concentrationScaled} and the evolution of the viscoelastic stress (Eqs. \eqref{eq:trSscaled} and \eqref{eq:devSscaled}). 
The time discretization of the concentration equation is done implicitly. The discretization of the surface stress equations is realised by an IMEX method, their exact scheme is given by Eqs. \eqref{eq:trSweak} and \eqref{eq:devSweak} below. 

If the surface is viscous, the surface force $\nabla_\Gamma \cdot S$ involves a second order derivative of the velocity $\vv{v}$. The explicit coupling of surface stress evolution to the Stokes system by Eq.~\eqref{eq:jumpConditionScaled} results in a harsh time step restriction. To circumvent this we use forward relaxation. Assume that $\vv{f}^n:=\nabla_\Gamma \cdot (S^n+S_a^n)+\vv{f}_c$ is the sum of the surface forces at time step $n$. Then we compute a relaxed surface force $\vv{F}^n$ by
\begin{equation}
    \vv{F}^n = \omega \vv{f}^n + (1-\omega) \vv{F}^{n-1}.
    \label{eq:forwardRelaxation}
\end{equation}
Here $\omega \in (0,1]$ and $\vv{F}^0 = 0$. 
This force $\vv{F}^n$ is then used to replace the surface force $\vv{f}$ in Eq.~\eqref{eq:jumpConditionScaled}. 
This forward relaxation is only needed for more viscous cases, so we chose to have $\omega = 0.01$ if $\max \{\frac{\hat{\eta_S}}{
\hat{\tau}_S}, \frac{\hat{\eta}_A}{\hat{\tau}_A} \} > 100$. In other cases, $\omega = 1$, and there is no forward relaxation. Even though only the viscoelastic force component causes instabilities and requires the relaxation, it is also applied to the other surface forces to ensure they are of similar magnitudes.

To sum up, each time step $n$ is constructed as follows: 
\begin{enumerate}
    \item Compute the velocity $\vv{v}^n$ from the Stokes equation using the active and viscoelastic surface forces $\vv{F}^n$ computed in the previous time step in the boundary condition (using the weak forms defined below in Eqs. \eqref{eq:stokes1weak} and \eqref{eq:stokes2weak}). 
    \item Use the computed velocity at $\Gamma$, to solve the equations for the surface concentration $c^n$ and the viscoelastic surface stress components $\bar{S}^n$ and ${\rm tr}(S^n)$ (using the weak forms defined below in Eqs. \eqref{eq:concentrationWeak}, \eqref{eq:trSweak} and \eqref{eq:devSweak}).
    \item Subtract a possibly small numerical error $\frac{\tr \bar{S}^n}{2} P$ from $\bar{S}^n$ to make it traceless again. Then calculate $S^n=\bar{S}^n+\frac{\tr(S^n)}{2} P^n$, $\nabla_\Gamma \cdot S^n$ and the active surface tension force. Use the forward relaxation defined in Eq. \eqref{eq:forwardRelaxation} to get the force $\vv{F}^n$ that will be imposed on the surface.
    \item Calculate the grid velocity $\vv{w}^n$ as a harmonic extension of the surface velocity grid velocity $\vv{w}_{|\Gamma}^n$, and move the meshes of $\Gamma$ and $\Omega$ accordingly. The surface grid velocity is the component of the velocity that is normal to the surface, $\vv{w}_{|\Gamma}^n = (\vv{n} \cdot \vv{v}^n_{|\Gamma}) \vv{n} + P\vv{v}_{avg}$. Here $\vv{v}_{avg}$ is the average velocity of the cell.
\end{enumerate}

\subsection{Spatial discretization}
Because of the rotational symmetry of the problem we use an axi-symmetric model as is illustrated in Fig. \ref{fig:domain} right. Instead of Cartesian coordinates we use the coordinates $z$, $r$ and $\phi$: $z$ for the location along the symmetry axis, $r$ for the distance from the symmetry axis and $\phi$ for the azimuthal position. Because of the rotational symmetry the variables do not change when changing $\phi$. Note that this does not imply that $v_\phi=0$, merely that $\partial_\phi f = 0$ for any function $f$ on $\Gamma$. 

For the spatial discretization we use the C++ finite element library AMDiS \cite{Vey2007, Witkowski2015}. The grid contains three meshes, two 2-dimensional meshes for the fluid and one 1-dimensional mesh for the surface $\Gamma$. The discretization for the fluid domains $\Omega_0$ and $\Omega_1$ will be referred to as $T_0$ and $T_1$. The mesh of the surface is named $T_\Gamma$. 
For the definitions of the differential operators in cylindrical coordinates in this section we used appendix D in \cite{Lautrup2004}.
To make the weak forms of the equations more readable we introduce the following notation. We define $\nabla'$ as the 2-dimensional gradient $(\partial_z, \partial_r)^T$. Similarly we define the velocity $\vv{v}' := (v_z, v_r) := (\hat{\vv{e}}_z\cdot\vv{v},\hat{\vv{e}}_r\cdot\vv{v})$, where $\hat{\vv{e}}_z$, $\hat{\vv{e}}_r$ are the base vectors of the cylindrical coordinate system (see Fig.~\ref{fig:domain}). 
The grid velocity of the surface $\vv{w}$ is only defined in the $z,r$-plane. 

For the Stokes Eqs. \eqref{eq:stokes1scaled} and \eqref{eq:stokes2scaled} we use the second order polynomial space for the velocity and an extended first order polynomial space for the pressure. The extension is required to allow for discontinuities of the pressure across the surface $\Gamma$ \cite{mokbel2020ale}. The finite elements spaces are then
\begin{equation}
    P_{i} = \Big\{ q \in C^0(\bar{\Omega}_i) \cap L^2(\Omega_i) \Big| \evalat[\big]{q}{k} \in P_1(k), k \in T_i \Big\} \text{ for } i=0,1     \label{eq:pFEMspace}
\end{equation}
for the pressure and 
\begin{equation}
    V = \Big\{ {u} \in C^0(\bar{\Omega}) \cap H^1(\Omega) \Big| \evalat[\big]{{v}}{k} \in P_2(k), k \in T_0 \cup T_1 \Big\} \label{eq:vFEMspace}
\end{equation}
for the components of velocity. Here $P_i(k)$ is the set polynomials of order $i$ on a domain $k$. This is an extension of the Taylor-Hood finite element space, which we choose for its optimal convergence for these low order elements. The weak formulation for the axi-symmetric version of Eqs. \eqref{eq:stokes1scaled} and \eqref{eq:stokes2scaled} then reads:\\
Find $(\vv{v}, p) \in V^3 \times P_0 \cup P_1$ such that for all $(\vv{u}, q) \in V^3 \times P_0 \cup P_1$ the following holds
\begin{equation}
\int_{\Omega_i} \left(\nabla' \cdot \vv{v}' + \frac{v_r}{r} \right) q d\Omega = V_{restore}, \label{eq:stokes1weak}
\end{equation}
\begin{align}
\begin{split}
\int_{\Gamma} \vv{F} \cdot \vv{u} d\Gamma
= \sum_{i=0}^1 \int_{\Omega_i}&-p \left(\nabla' \cdot \vv{u}' \right) +\frac{\eta_i}{\eta_1} \left((\nabla'\vv{v}' + (\nabla'\vv{v}')^T) : \nabla'\vv{u}' \right) \\ 
&+ \frac{\eta_i}{\eta_1} \left( \nabla'v_\phi \cdot \nabla'u_\phi +\frac{2v_r u_r + 2v_\phi u_\phi}{r^2} \right) \\
&- \frac{\eta_i}{\eta_1} \left( \frac{1}{r}\left(u_z(\partial_z v_r  + \partial_r v_z)+  2u_r\partial_r v_r +2u_\phi \partial_r v_\phi + v_\phi \partial_r u_\phi\right) \right) d\Omega
\end{split}
\label{eq:stokes2weak}
\end{align}
Again, $\vv{F}$ is the relaxed sum of the surface forces (Eq.~\eqref{eq:forwardRelaxation}). Moreover, $V_{restore}$ is a scalar that helps conserve the volume of the inner domain, which due to numerical errors is not perfectly conserved. It is defined as $V_{restore} = C_{restore} \frac{V(t=0) - V(t)}{V(t)}$. Here $V(t)$ is the volume of the inner domain $\Omega_1$ at time $t$ and $C_{restore}$ is a constant which we choose to be $\frac{0.125}{\Delta t}$.

For the discretisation of the viscoelastic equations in cylindrical coordinates we first redefine the rows and columns of the tensors. For a tensor $S$ instead of $x$, $y$ and $z$ we now use 
\begin{equation}
    S = \begin{pmatrix} S_{zz} & S_{zr} & S_{z \phi} \\ S_{rz} & S_{rr} & S_{r \phi} \\ S_{\phi z} & S_{\phi r} & S_{\phi \phi} \end{pmatrix}
\end{equation}
where $S_{ij} = \hat{\vv{e}}_i \cdot S \cdot \hat{\vv{e}}_j$ for $i,j\in \{z,r,\phi\}$. The gradient of the velocity in cylindrical coordinates is defined as \cite{Lautrup2004}
\begin{equation}
\nabla \vv{v} = \begin{pmatrix} (\nabla'\vv{v}') & \begin{matrix} 0 \\ -v_\phi/r \end{matrix} \\ \begin{matrix} \partial_z v_{\phi} & \partial_r v_{\phi} \end{matrix} & v_r/r \end{pmatrix}.
\label{eq:grad_v}
\end{equation}
The normal $\vv{n}$ in cylindrical coordinates at $\phi = 0$ is defined by 
\begin{equation}
\vv{n} = \begin{pmatrix} n_z \\ n_r \\ 0 \end{pmatrix},
\label{eq:n_rotational}
\end{equation}
hence the projection matrix $P$ is defined by 
\begin{equation}
P = \begin{pmatrix} P' & \begin{matrix} 0 \\ 0 \end{matrix} \\ \begin{matrix}  0 & 0\end{matrix} & 1\end{pmatrix},
\label{eq:P_rotational}
\end{equation}
where $P_{ij}' = \hat{\vv{e}}_i\cdot P \hat{\vv{e}}_j$ with $i,j \in \{z,r\}$.
To compute the viscoelastic stress we use the definitions in  Eqs. \eqref{eq:grad_v} and \eqref{eq:P_rotational} and substitute them in Eqs. \eqref{eq:devSscaled} and \eqref{eq:trSscaled}. 
The finite element space for the surface stress, $\mathcal{S}_\Gamma$, is defined by first order polynomials, 
\begin{equation}
    \mathcal{S}_\Gamma = \Big\{ \psi \in C(\Gamma) \cap L^2(\Gamma) \Big| \evalat[\big]{\psi}{k} \in P_1(k), k \in T_\Gamma \Big\}.
\end{equation}
The weak form is then: Find $\tr S^{n+1} \in \mathcal{S}_\Gamma$ and $\bar{S}_{ij} \in \mathcal{S}_\Gamma$ such that for all $\psi,\psi_{ij} \in \mathcal{S}_\Gamma$ and $i,j \in \{z, r, \phi\}$ the following equations hold, 
\begin{align}
    &\int_{\Gamma} \tr S^{n+1}\left(1 + \frac{\hat{\tau}_A}{\Delta t}\right) \psi d \Gamma \nonumber\\
    &~~~= \int_{\Gamma} \left( 2 \tr D + \hat{\tau}_A\left( \frac{2\hat{\eta}_S}{\hat{\eta}_A}(\bar{S}^n:\nabla_\Gamma \vv{v}) + \tr (S^n)\tr (D) - \left((\vv{v}'-\vv{w})\cdot \nabla_\Gamma'  \right) \tr S^{n} + \frac{\tr S^n}{\Delta t} \right) \right) \psi d\Gamma,
\label{eq:trSweak} \\
    &\int_{\Gamma} \bar{S}_{ij}^{n+1}\left(1 + \frac{\hat{\tau}_S}{\Delta t}\right) \psi_{ij}  d \Gamma \nonumber\\
    &~~~= \int_{\Gamma} \left( P \left( 2 \bar{D} - \hat{\tau}_S\left(\left((\vv{v}'-\vv{w})\cdot \nabla_\Gamma' \right) \bar{S}^n -  \frac{\bar{S}^n}{\Delta t} - \nabla_\Gamma \vv{v} \bar{S}^n - (\nabla_\Gamma \vv{v})^T \bar{S}^n \right) \right) P \right)_{ij}\psi_{ij} d\Gamma \label{eq:devSweak}\\
    &~~~~~~- \int_{\Gamma} \hat{\tau}_S \left( P(\bar{S}:\nabla_\Gamma \vv{v}^n) - \frac{\hat{\eta}_A}{\hat{\eta}_S}\tr(S^n) \bar{D}  \right)_{ij}\psi_{ij} d\Gamma. \nonumber
\end{align}
Here $\Gamma_n$ is the surface at time $t=t_n$ and $\Delta t$ is the time step size. The vectors and tensors are in cylindrical coordinates. The surface gradient and surface rate of deformation are defined as $\nabla_\Gamma \vv{v} = \nabla \vv{v} P$ and $P \frac{1}{2}\left( \nabla \vv{v} + (\nabla \vv{v})^T \right) P$ respectively, with $\nabla \vv{v}$ and $P$ defined as in Eqs. \eqref{eq:grad_v} and \eqref{eq:P_rotational}.

To calculate the surface divergence of the stress tensor (Eq. \eqref{eq:jumpConditionScaled}) in cylindrical coordinates, let us first consider a continuous extension of the surface tensor $S$ to the fluid. We name this extension $T$. The surface divergence $\nabla_\Gamma \cdot S$ is then defined as $\nabla T \cdot P$. The gradient of the tensor in cylindrical coordinates is
\begin{equation}
\begin{aligned}
\nabla T &=\frac{\partial T_{z r}}{\partial r} \hat{\vv{e}}_z \otimes \hat{\vv{e}}_r \otimes \hat{\vv{e}}_r+\frac{1}{r}\left(\frac{\partial T_{z r}}{\partial \phi}-T_{z \phi}\right) \hat{\vv{e}}_z \otimes \hat{\vv{e}}_r \otimes \hat{\vv{e}}_\phi+\frac{\partial T_{z r}}{\partial z} \hat{\vv{e}}_z \otimes \hat{\vv{e}}_r \otimes \hat{\vv{e}}_z \\
&+\frac{\partial T_{z \phi}}{\partial r} \hat{\vv{e}}_z \otimes \hat{\vv{e}}_\phi \otimes \hat{\vv{e}}_r+\frac{1}{r}\left(\frac{\partial T_{z \phi}}{\partial \phi}+T_{z r}\right) \hat{\vv{e}}_z \otimes \hat{\vv{e}}_\phi \otimes \hat{\vv{e}}_\phi+\frac{\partial T_{z \phi}}{\partial z} \hat{\vv{e}}_z \otimes \hat{\vv{e}}_\phi \otimes \hat{\vv{e}}_z \\
&+\frac{\partial T_{z z}}{\partial r} \hat{\vv{e}}_z \otimes \hat{\vv{e}}_z \otimes \hat{\vv{e}}_r+\frac{1}{r} \frac{\partial T_{z z}}{\partial \phi} \hat{\vv{e}}_z \otimes \hat{\vv{e}}_z \otimes \hat{\vv{e}}_\phi+\frac{\partial T_{z z}}{\partial z} \hat{\vv{e}}_z \otimes \hat{\vv{e}}_z \otimes \hat{\vv{e}}_z \\
&+\frac{\partial T_{r r}}{\partial r} \hat{\vv{e}}_r \otimes \hat{\vv{e}}_r \otimes \hat{\vv{e}}_r+\frac{1}{r}\left(\frac{\partial T_{r r}}{\partial \phi}-\left(T_{\phi r}+T_{r \phi}\right)\right) \hat{\vv{e}}_r \otimes \hat{\vv{e}}_r \otimes \hat{\vv{e}}_\phi+\frac{\partial T_{r r}}{\partial z} \hat{\vv{e}}_r \otimes \hat{\vv{e}}_r \otimes \hat{\vv{e}}_z \\
&+\frac{\partial T_{r \phi}}{\partial r} \hat{\vv{e}}_r \otimes \hat{\vv{e}}_\phi \otimes \hat{\vv{e}}_r+\frac{1}{r}\left(\frac{\partial T_{r \phi}}{\partial \phi}+\left(T_{r r}-T_{\phi \phi}\right)\right) \hat{\vv{e}}_r \otimes \hat{\vv{e}}_\phi \otimes \hat{\vv{e}}_\phi+\frac{\partial T_{r \phi}}{\partial z} \hat{\vv{e}}_r \otimes \hat{\vv{e}}_\phi \otimes \hat{\vv{e}}_z \\
&+\frac{\partial T_{r z}}{\partial r} \hat{\vv{e}}_r \otimes \hat{\vv{e}}_z \otimes \hat{\vv{e}}_r+\frac{1}{r}\left(\frac{\partial T_{r z}}{\partial \phi}-T_{\phi z}\right) \hat{\vv{e}}_r \otimes \hat{\vv{e}}_z \otimes \hat{\vv{e}}_\phi+\frac{\partial T_{r z}}{\partial z} \hat{\vv{e}}_r \otimes \hat{\vv{e}}_z \otimes \hat{\vv{e}}_z \\
&+\frac{\partial T_{\phi r}}{\partial r} \hat{\vv{e}}_\phi \otimes \hat{\vv{e}}_r \otimes \hat{\vv{e}}_r+\frac{1}{r}\left(\frac{\partial T_{\phi r}}{\partial \phi}+\left(T_{r r}-T_{\phi \phi}\right)\right) \hat{\vv{e}}_\phi \otimes \hat{\vv{e}}_r \otimes \hat{\vv{e}}_\phi+\frac{\partial T_{\phi r}}{\partial z} \hat{\vv{e}}_\phi \otimes \hat{\vv{e}}_r \otimes \hat{\vv{e}}_z \\
&+\frac{\partial T_{\phi \phi}}{\partial r} \hat{\vv{e}}_\phi \otimes \hat{\vv{e}}_\phi \otimes \hat{\vv{e}}_r+\frac{1}{r}\left(\frac{\partial T_{\phi \phi}}{\partial \phi}+\left(T_{r \phi}+T_{\phi r}\right)\right) \hat{\vv{e}}_\phi \otimes \hat{\vv{e}}_\phi \otimes \hat{\vv{e}}_\phi+\frac{\partial T_{\phi \phi}}{\partial z} \hat{\vv{e}}_\phi \otimes \hat{\vv{e}}_\phi \otimes \hat{\vv{e}}_z \\
&+\frac{\partial T_{\phi z}}{\partial r} \hat{\vv{e}}_\phi \otimes \hat{\vv{e}}_z \otimes \hat{\vv{e}}_r+\frac{1}{r}\left(\frac{\partial T_{\phi z}}{\partial \phi}+T_{r z}\right) \hat{\vv{e}}_\phi \otimes \hat{\vv{e}}_z \otimes \hat{\vv{e}}_\phi+\frac{\partial T_{\phi z}}{\partial z} \hat{\vv{e}}_\phi \otimes \hat{\vv{e}}_z \otimes \hat{\vv{e}}_z.
\end{aligned}
\end{equation}
All derivatives w.r.t. $\phi$ are zero because of axi-symmetry. If we then take the product with the projection matrix $P$ as defined in Eq. \eqref{eq:P_rotational} we get the surface divergence,
\begin{equation}
    \nabla T \cdot P = \begin{pmatrix} P_{zz} \pdv{T_{zz}}{z} + P_{zr}\pdv{T_{zz}}{r} + P{rz}\pdv{T_{zr}}{z} + P_{rr}\pdv{T_{zr}}{r} + \frac{1}{r} T_{zr}\\
    P_{rr}\pdv{T_{rr}}{r} + P_{rz}\pdv{T_{rr}}{z} + P_{zr}\pdv{T_{rz}}{r} + P_{zz} \pdv{T_{rz}}{z} + \frac{1}{r}(T_{rr} - S_{\phi \phi})\\
    P_{rr}\pdv{T_{\phi r}}{r} + P_{rz}\pdv{T_{\phi r}}{z} + P_{zr} \pdv{T_{\phi z}}{r} + P_{zz}\pdv{T_{\phi z}}{z} + \frac{1}{r}(T_{r \phi} + T_{\phi r})
    \end{pmatrix}.
\end{equation}


Finally, for the finite element space of the surface concentration we use second order polynomials, 
\begin{equation}
    C_\Gamma = \Big\{ \psi \in C(\Gamma) \cap L^2(\Gamma) \Big| \evalat[\big]{\psi}{k} \in P_2(k), k \in T_\Gamma \Big\}.
\end{equation} 
We adopt the weak form for the surface concentration from \cite{Lucas_paper}: find $c^{n+1} \in C_\Gamma$ such that for all $\psi \in C_\Gamma$ the following holds
\begin{equation}
    0 = \int_{\Gamma} \left( \frac{c^{n+1} - c^n}{\Delta t} - c^{n+1}(\vv{v}' - \vv{w}) \cdot \nabla_\Gamma' \psi + c^{n+1} \psi \nabla_\Gamma' \cdot \vv{w} + \hat{D}_c \nabla_\Gamma' c^{n+1} \cdot \nabla_\Gamma' \psi \right)r d\Gamma. \label{eq:concentrationWeak}
\end{equation}

\subsection{Verification of the numerical scheme}
To test the convergence of the spatial and time discretization, we choose the parameters $\hat{\tau}_A = \hat{\tau}_S = 100$ and $\hat{\eta}_A = \hat{\eta}_S = 10$ as a test case. This case is chosen because it displays significant deformations in $\Gamma$. To test the convergence we compare the equatorial radius $r$ for various grid and time step sizes.

For the time steps we use $\Delta t = 10^{-4}, 0.5\cdot 10^{-4}, 0.25\cdot 10^{-4}, 0.125\cdot 10^{-4}$. Using Richardson extrapolation we determine the order of convergence to be 1, as would be expected from the backward and forward Euler schemes that are used. The difference between the solutions is of the order $10^{-6}$. 

To check the spatial discretization error, we use different grid distances ($0.08, 0.04, 0.02, 0.01$) on the surface and its vicinity. Using Richardson extrapolation we determine the order of convergence to be 1, which is expected for the first order polynomials used in Eqs. \eqref{eq:trSweak} and \eqref{eq:devSweak}. The difference between the solutions is of the order $10^{-3}$ between the solutions with a grid density of $0.02$ and $0.01$. We conclude that a grid density of $0.02$ and a time step size $\Delta t = 10^{-4}$ are fine enough for sufficiently accurate solutions and used these values throughout this paper.

\section{Results}

\subsection{Comparison of the polar and ring modes}
\label{app:correlation}
The polar and ring mode are the first and second mode in spherical harmonics if only rotationally symmetric modes are considered. The first and second mode are defined by $g_1(\theta) = \cos(\theta)$ and $g_2(\theta) = \frac{1}{2} (3 \cos^2(\theta) -1)$ respectively. $\theta 
\in [0, \pi]$ is the angle w.r.t. $z$-axis. The comparison of these modes in the simulations is done by calculating the correlation coefficient of the solution $c$ and the ring and polar modes, given by
\begin{equation}
    \rho_l = \frac{\int_\Gamma (c - \overline{c}) g_l d \Gamma}{\sqrt{\int_\Gamma g_l^2 d \Gamma \int_\Gamma (c - \overline{c})^2 d \Gamma}}.
    \label{eq:corrCoeff}
\end{equation}
We compare $\rho_1$ and $\rho_2$ to study the ring slipping in Sec. \ref{sec:contractileRing}. If a ring slipped $\rho_1$ would go to $1$ or $-1$ and $\rho_2$ would go from a value close to $-1$ to $0$.

Equation \eqref{eq:corrCoeff} only holds if the surface is a sphere. For this we define a measure $\mu(\Gamma)$ to calculate the deformation from a sphere. The correlation coefficient is then only used if $\mu(\Gamma) < 0.01$. The measure is defined as
\[
\mu(\Gamma) = \int_\Gamma \left(\frac{1}{2} \kappa - \frac{1}{R_s}\right)^2 d\Gamma,
\]
where $R_s$ is the radius of a sphere with the same area as surface $\Gamma$, and $\kappa$ is the mean curvature.


\subsection{Viscous force}
\label{app:viscousForce}
Here, we show that a purely viscous spherical surface does not exhibit any shape changes if exposed to a counter rotating force field. 

If we assume the initial surface to be a sphere and assume that the viscous force of the surrounding fluids is negligible, then we can use spherical coordinates to express the dynamics. The generic chiral force from Eq.~\eqref{eq:chiralForce} in spherical coordinates becomes 
\[
\vv{f} = \begin{pmatrix} 0 \\ 0\\ \sin(\phi) \cos(\phi) \end{pmatrix} = \begin{pmatrix} 0 \\ 0\\ \frac{\sin(2\phi)}{2} \end{pmatrix}
\]
and the force balance from Eq. \eqref{eq:jumpConditionScaled} becomes $\vv{f} + \nabla \cdot S = 0$.
Let us assume that the velocity $\vv{v}$ is proportional to $\vv{f}$, i.e. $\vv{v} = \alpha \vv{f}$ for some $\alpha \in \mathbb{R}$. 
Clearly this velocity field does not lead to any changes in surface area, $\nabla_\Gamma \cdot \vv{v}=0$. 
Now we can calculate the rate of deformation $D = \bar{D} = \frac{1}{2}\left(\nabla_\Gamma \vv{v} + (\nabla_\Gamma \vv{v}) \right)$. For the viscous case, the stress $S$ reduces to $S = 2\eta_S D$. From this, we can calculate the viscous force $\vv{f}_{\text{viscous}} = \nabla \cdot S$ and obtain
\[
\vv{f}_{\text{viscous}} = \alpha \begin{pmatrix} 0 \\ 0 \\ -3 \eta_S \sin (2\phi) \end{pmatrix},
\]
Now we see that that for $\alpha = \frac{1}{6\eta_S}$ the force balance $\vv{f} + \nabla \cdot S = 0$ holds. So the solution for the velocity is $\vv{v} = \frac{1}{6\eta_S} \vv{f}$ which is parallel to the surface and hence does not change its shape.

\subsection{Parameters relevant for biological cells}
\label{app:biology}
The following parameters have been estimated for biological cells:
\begin{itemize}
    \item The radius of the cell, $R = 10\,\mu$m.
    \item The interior of the cell, i.e. the cytoplasm, was measured to have viscosities with values between $1-10$~Pa s \cite{dani06}. 
    \item The areal viscosity of the surface was previously estimated as $\eta_A \in [0.01, 0.1]$~Ns/m.  \cite{Fischer-Friedrich2016, Hosseini2021}.
    \item The shear viscosity of the surface, $\eta_S$ can be approximated as $\frac{1}{3} \eta_A$ corresponding to a Poisson ratio of 0.5 \cite{Mokbel2020}.
    \item The relaxation times $\tau_S$ and $\tau_A$ are approximately $60$ seconds according to measured times of molecular turnover \cite{Salbreux2012}.
    \item Naganathan {\it et al.}\cite{Naganathan2014} found a chiral velocity $v_{bio} \in [10^{-8}, 10^{-7}]$~m/s. 
\end{itemize}
Using the above estimates, we associate the following ranges of scaled parameters:
\begin{itemize}
    \item The areal surface viscosity $\hat{\eta}_A \in [10^2, 10^4]$ using as surface reference viscosity $R\eta_1$ with $\eta_1$ between $1-10\,$Pa~s.
    \item The surface shear viscosity $\hat{\eta}_S \in [\frac{1}{3} \cdot 10^2, \frac{1}{3} \cdot 10^4]$.
    \item In our simulations, we observe maximal chiral dimensionless velocities $v_{num}$ between $10^{-3}$ and $10^{-1}$. We calculate the range for the velocity scale $v_{sc} \in [10^{-7}, 10^{-4}]$~m/s using $v_{bio} =v_{sc} v_{num}$. Dividing the length scale $R$ by the velocity scale gives us the timescale which we use to derive the range for the dimensionless relaxation times $\hat{\tau}_A, \hat{\tau}_S \in [0.6, 600]$.
\end{itemize}
The parameters used in the parameter screens are $\hat{\eta}_A, \hat{\eta}_S \in [0.1, 10^4]$, $\hat{\tau}_S, \hat{\tau}_A \in [0.01, 1000]$ and $\hat\eta_1=\hat\eta_0$ except for simulations corresponding to Fig.~\ref{fig:ringSlip} where $\hat\eta_1=10\hat\eta_0$. 
The ranges of scaled parameters inferred from measurements in biological cells are very broad, but they are contained in the parameter ranges chosen in our simulations.
\newpage

\printbibliography
\end{document}